\documentclass[useAMS,usenatbib]{mn2e}
\usepackage{graphics,graphicx,epsfig,psfig}
\usepackage[normalem]{ulem}
\usepackage[dvipsnames]{color}
\usepackage{soul,xcolor}
\usepackage{amsmath, amssymb}
\usepackage[]{inputenc,amssymb}
\usepackage{upgreek,textgreek}
\usepackage{booktabs,caption}
\usepackage{amsmath}
\usepackage{multirow}

\setstcolor{red}

\def \be{\begin{equation}}
\def \ee{\end{equation}}
\def \bea{\begin{eqnarray}}
\def \eea{\end{eqnarray}}
\def \etal{{et al.}}

\definecolor{webgreen}{rgb}{0,.5,0}
\definecolor{webbrown}{rgb}{.6,0,0}

\usepackage[%
    colorlinks = true,%
    linkcolor = blue,%
    urlcolor  = blue,%
    citecolor = webgreen,%
    anchorcolor = blue]{hyperref}

\newcommand{\ufhref}[3][blue]{\href{#2}{\color{#1}{#3}}}%

\title[Cosmic rays from compact young star clusters]{Constraining cosmic ray acceleration in young star clusters using multi-wavelength observations}
\author[Gupta, Nath \& Sharma]
{Siddhartha Gupta$^{1,2}$ \thanks{E-mail: siddhartha@rri.res.in}, Biman B. Nath$^1$, Prateek Sharma$^2$\\
$^1$Raman Research Institute, Sadashiva Nagar, Bangalore 560080, India\\
$^2$Joint Astronomy Programme and Department of Physics, Indian Institute of Science, Bangalore 560012, India}
\begin{document}
\maketitle
\label{firstpage}
\begin{abstract}
We use $1$D and $3$D two-fluid cosmic ray (CR) hydrodynamic simulations to investigate the role of CRs in the vicinity of a compact young star cluster. We model a self-gravitating cloud (density profile $\rho \propto r^{-1}$), include important thermal and non-thermal processes, and explore two different CR injection scenarios. We show that if internal shocks in the wind-driving region are the main site for CR acceleration, then the resulting $\gamma$-ray luminosity ($L_{\rm \gamma}$) can reach $\approx 5\%$ of the mechanical luminosity ($L_{\rm w}$), independent of the fraction of wind energy ($\sim 1-20\%$) injected into CRs. In contrast, if the forward/reverse shock of a bubble is the injection site then $L_{\rm \gamma}$ increases linearly with the CR injection fraction, as expected analytically. We find that the X-ray luminosity ($L_{\rm x}$) in the forward/reverse shock injection scenario is $\gtrsim 10^{-3} L_{\rm w}$, which is $\sim 10$ times larger than in the central wind-driving injection case. We predict the corresponding range of the synchrotron radio luminosity. We show how multi-wavelength observations can constrain the CR parameters. Comparing the predicted multi-wavelength luminosities with those of 30 Doradus we identify the reverse shock as the most probable CR injection site, and that thermal conduction is important. We do not find significant dynamical impact of CRs in our models.
\end{abstract}
\section{Introduction} \label{sec:intro}
Star clusters are among the most fundamental objects in a galaxy. They are located in the core of dense molecular clouds and contain several thousand solar mass (for a review see \citealt{Longmore2014}). The stars energize the surrounding medium, leading to gas expulsion and the formation of interstellar bubbles (ISBs).

The theoretical modeling of ISBs serves as a standard scenario for the wind and ISM interaction (\citealt{Weaver1977}). Observations in X-rays, ultraviolet and infrared (e.g., \citealt{Chu2003a}; \citealt{Townsley2006}, \citealt{Lopez2014}) have helped in our understanding of ISBs. Recent works have attempted to relax some of the assumptions in the standard scenario, for example, include the effect of different forms of pressure other than thermal pressure, or include the effect of spatial distribution of stars. It has been found that the dynamics of ISBs strongly depend on the clustering of stars and on the ambient density (e.g., \citealt{Nath2013}; \citealt{Krause2013}; \citealt{Sharma2014}; \citealt{Kim2015}; \citealt{Martizzi2015}; \citealt{Yadav2017}, \citealt{Vasiliev2017}). The effect of stellar radiation  has also been studied (\citealt{Harper2009}; \citealt{Silich2013}; \citealt{Dale2013}). It has been shown that radiation pressure can boost gas expulsion in the early phase ($\lesssim 1$ Myr) whereas the late time evolution is governed by the mechanical energy injection and photo-heating (\citealt{Gupta2016}). There is another promising driving mechanism, namely, the pressure due to relativistic particles such as cosmic rays (CRs), whose effects are yet to be understood in detail.

Star forming regions have been thought to be efficient sites for CR acceleration (\citealt{Knodlseder2013}; \citealt{Bykov2014}; \citealt{Aharonian2018}). Several ISBs have been identified as powerful sources of gamma-rays (hereafter, $\gamma$-rays) [The Fermi and H.E.S.S. collaboration]. \citet{Ackermann2011} found that the Cygnus OB association is quite bright in GeV range. \citet{Yang2018} reported $\gamma$-ray emission in Westerlund $2$. High energy photons have also been detected from the Large Magellanic Cloud(LMC). It has been reported that a massive star cluster, 30 Doradus,  produces both GeV and TeV photons (\citealt{Abdo2010}, \citealt{Abramowski2015}). In a few cases, the $\gamma$-ray luminosity is $\sim 1\%$ of the wind mechanical power, and it is almost comparable to the X-ray luminosity (c.f. Table \ref{tab:obs_data}). Furthermore, \citet{Hughes2004} concluded that 30 Doradus dominates the radio continuum emission in LMC at $1.4$ GHz (see also \citealt{Murphy2012}; \citealt{Foreman2015}). These emissions occur when relativistic particles interact with the magnetic field and matter, and confirm the presence of CRs in ISBs. It is then reasonable to ask to what extent CRs affect the dynamics and evolution of ISBs.

There is yet another motivation to study the effect of CRs on ISBs. At a larger length-scale, it has been suggested that CRs can dynamically affect galactic winds (\citealt{Booth2013}; \citealt{Salem2013}; \citealt{Simpson2016}; \citealt{Wiener2017}). However, the detailed physics is not clearly understood. Firstly, changing the adiabatic index of the gas from $5/3$ to $4/3$ (i.e. replacing thermal pressure by CR pressure) {\it reduces} the size of ISBs (e.g. see Equation ($4$) in \citealt{Gupta2018}; also see \citealt{Chevalier1983}). Secondly, diffusion of CRs would tend to {\it decrease} the pressure gradient, and therefore {\it reduce} the dynamical effect of CRs. We propose to study these processes in an ISB, which may help us to understand the effects at a larger length-scale.  

In an earlier work, we studied the effect of CRs in an idealized ISB \citep{Gupta2018}. We found that the effect of CRs mainly depends on the CR injection region, diffusion coefficient and the shock Mach number. CRs can be injected in two different ways. In one case, CRs are injected at spatially resolved shocks whereas in the other case, it is assumed that a small fraction ($\sim 10\%$) of the wind/supernovae energy directly goes to CRs via internal shocks (these internal shocks may originate due to stellar flares, colliding winds and supernovae which are difficult to resolve in numerical simulations). The basic difference in (spatially  resolved) shock injection and central injection of CRs is that, in the latter case, the back reaction from CRs at the shock can modify the thermodynamic properties of the shock when the Mach number exceeds $\gtrsim 12$ (\citealt{Drury1981}; \citealt{Drury1986}; \citealt{Becker2001}). In this case most of the upstream kinetic energy goes into CRs. This is how diffusive shock acceleration is captured in a two-fluid model. We estimated various relevant time-scales for the CR affected bubbles (see sections $2.2$, $4.2$ in \citealt{Gupta2018}). We showed that CR dominated ISBs may contain comparatively cool thermal plasma (temperature $\sim 10^{6.5}$ K), even in the absence of thermal conduction (which can also reduce the interior temperature of an ISB).

In this paper, we extend our work to determine the multi-wavelength signatures of ISBs arising from the presence of CRs, with the help of 1D and 3D numerical simulations. This will help us to compare our findings with observations of ISBs in different wavelengths, and to constrain the CR injection parameters. 

We focus on the early evolution ($\lesssim 4$ Myr) when mechanical wind from a compact star cluster can form a reverse (termination) shock. We do not include supernova explosion (e.g. \citealt{Sharma2014}; \citealt{Kim2015}; \citealt{Yadav2017}; \citealt{Vasiliev2017}) or large spatial separation of stars, which may change the evolution and structure of the ISBs. We start with an analytic estimates of different luminosites for a two-fluid ISB in \S \ref{sec:analy}. In \S \ref{sec:obs} we discuss some recent results from ISB observations. This helps us to set-up our simulation, as discussed in \S \ref{sec:sim_setup}. The results are presented in \S \ref{sec:result} and \S \ref{sec:discussion}, and summarized in \S \ref{sec:summary}.
\section{Analytical estimates} \label{sec:analy}
We consider an idealized two-fluid model of an ISB (for details, see \citealt{Gupta2018}). We wish to estimate the $\gamma$-ray, X-ray and radio luminosities, considering that CRs are being accelerated in an ISB.
\subsection{$\gamma$-ray} \label{subsec:gr}
The nature of $\gamma$-ray emission depends on the interaction mechanism between CRs and matter \citep{Mannheim1994}. 
\subsubsection{Hadronic origin} \label{subsubsec:hg}
To estimate $\gamma$-ray luminosity due to hadronic interaction, we use the analytical prescription of \citet{Pfrommer2004}, which is briefly discussed below.

The $\gamma$-ray luminosity in $({E_{\rm \gamma 1}}-{E_{\rm \gamma 2}})$ energy band can be estimated using
\begin{eqnarray}
\label{eq:q_gamma}
L^{\rm H}_{\rm \gamma} & = & \int_{V}dV \int_{E_{\rm \gamma 1}}^{E_{\rm \gamma 2}}dE_{\rm \gamma} \, E_{\rm \gamma}\, q_{\rm \gamma}(n_{\rm N},e_{\rm cr}, E_{\rm \gamma}) \\ \nonumber 
& = &  \Delta V \, n_{\rm N}\,  e_{\rm cr}\, \left[\int_{E_{\rm \gamma 1}}^{E_{\rm \gamma 2}}dE_{\rm \gamma} \, E_{\rm \gamma}\, \tilde{q}_{\rm \gamma}(E_{\rm \gamma}) \right]  ,
\end{eqnarray}
where $q_{\rm \gamma}= dN/(dt\,dV\,dE_{\rm \gamma})$ is the number of $\gamma$-ray photons emitted per unit volume per unit time per unit energy, which is proportional to $n_{\rm N}$  (the number density of target nucleon) and $e_{\rm cr}$ (the CR energy density), and $\Delta V$ is the volume of the emitting region. The function $\tilde{q}_{\rm \gamma}$ is given as,
\begin{equation}
\tilde{q}_{\rm \gamma}=\left[\frac{ \sigma_{\rm pp} c\,\left(\frac{E_{\pi^{0}}}{\rm GeV}\right)^{-\alpha_{\rm \gamma}} \left[\left(\frac{2E_{\rm \gamma}}{E_{\rm \pi^{0}}}\right)^{\delta_{\rm \gamma}}+\left(\frac{2E_{\rm \gamma}}{E_{\rm \pi^{0}}}\right)^{-\delta_{\rm \gamma}}\right]^{-\alpha_{\gamma}/\delta_{\gamma}}}{ \xi^{\alpha_{\rm \gamma}-2} \left(\frac{3\alpha_{\rm \gamma}}{4}\right) \frac{E_{\rm p}}{2(\alpha_{\rm p}-1)}\left(\frac{E_{\rm p}}{\rm GeV}\right)^{1-\alpha_{\rm p}}{\it \beta}(\frac{\alpha_{\rm p}-2}{2},\frac{3-\alpha_{\rm p}}{2})}\right]\ .
\end{equation}
Here $E_{\rm p}/E_{\rm \pi^{0}}$ is the rest mass energy of proton/pions ($\pi^{\rm 0}$), $\alpha_{\rm p}$ and $\alpha_{\rm \gamma}$ are the spectral indices of the incident CR protons and emitted $\gamma$-ray photons respectively, $\delta_{\gamma} = 0.14\alpha_{\rm \gamma}^{-1.6} + 0.44$ is the spectral shape parameter and $ \sigma_{\rm pp}=32(0.96+e^{4.4-2.4\alpha_{\rm \gamma}})$ mbarn (see Equations (8), (19)-(21) in \citealt{Pfrommer2004}). 

From Equation (\ref{eq:q_gamma}), we find that the result of the integration from $0.1$ to $200$ GeV energy is $\approx 1.1\times 10^{-16}\,{\rm cm^{3}\,s^{-1}}$ and it depends weakly (error $< 20\%$) on the choice of $\alpha_{\rm p}$ or $\alpha_{\rm \gamma}$ ($2.1-2.5$) when $\alpha_{\rm \gamma}=\alpha_{\rm p}$ (e.g \citealt{Dermer1986}). The $\gamma$-ray spectrum beyond $200$ GeV differs from model to model, and we have, therefore, excluded it from our analysis. We thus obtain the $\gamma$-ray luminosity in $\approx (0.1-100)$ GeV band:
\begin{equation}
\label{eq:Lum_gamma}
L^{\rm H}_{\rm \gamma} \simeq 1.1\times 10^{-16} \left(\frac{\Delta V}{\rm cm^{3}}\right) \left(\frac{n_{\rm N}}{\rm cm^{-3}}\right)  \left(\frac{e_{\rm cr}}{\rm erg\,cm^{-3}}\right) \ {\rm erg\,s^{-1}}.
\end{equation}
Clearly $L_{\rm\gamma}$ is directly proportional to the target nucleon ($n_{\rm N}$) and the CR energy density ($e_{\rm cr}$), and therefore, the $\gamma$-ray emission arises from the denser region of the ISBs, e.g the swept-up ambient medium (shell). 

Consider the ambient density profile to be $\rho(r)=\rho_{\rm c}\left(r_{\rm c}/r\right)^{\rm s}\,$ where $\rho_{\rm c}/r_{\rm c}$ is the core density/radius of the ambient medium. We denote the CR pressure fraction in the shell as $W_{\rm sh} = P_{\rm cr}/(P_{\rm th}+P_{\rm cr})$ [$P_{\rm cr/th}$ is the volume averaged CR/thermal pressure in the shell]. From the self-similar evolution of the bubble we obtain
\begin{eqnarray} \label{eq:Lum_g_f}
L^{\rm H}_{\rm \gamma}= A\, W_{\rm sh}\ L_{\rm w}^{\rm (5-2s)/(5-s)}\,\left(\rho_{\rm c}r_{\rm c}^{\rm s}\right)^{\rm 5/(5-s)}\,  t_{\rm dyn}^{\rm (5-4s)/(5-s)}
\end{eqnarray}
where 
\begin{eqnarray} \label{eq:c1c2}
A & = & \frac{13.2\pi \times 10^{-16}}{m\rm{_H}} \left(\frac{21-6s}{(5-s)^2(3-s)^2}\right)\\  \nonumber
 & \times &\left[\frac{(\gamma-1)\,(5-s)^3\,(3-s)}{4 \pi \{(63-18 s)\gamma+s(2s+1)-28\}}\right]^{(5-2s)/(5-s)}
\end{eqnarray}
Here we have used Equations ($4$) and ($5$) in \citet{Gupta2018} to estimate the shell volume $\Delta V$ ($=4\pi R^{2} \Delta R$, $\Delta R$ is the shell width and $R$ is the radius of the ISB) and target density $n_{\rm N}$ ($\approx 4\times \rho(R)/m_{\rm H}$). We also have taken CR energy density $e_{\rm cr}=P_{\rm cr}/(\gamma_{\rm cr}-1)$ where $\gamma_{\rm cr}=4/3$.

Equation (\ref{eq:Lum_g_f}) shows that, for a fixed\footnote{Depending on CR injection model, $W_{\rm sh}$ may evolve with time, discussed in section \ref{subsec:gamma_sim}.} $W_{\rm sh}$, the time evolution of $\gamma$-ray luminosity depends on the ambient density power-law index `$s$'. If $5>s>5/4$, then $L^{\rm H}_{\rm \gamma}$ decreases with time. This is reasonable because the density falls so rapidly that only small column density targets are available for hadronic interaction. For $s<5/4$, $L^{\rm H}_{\rm \gamma}$ is an increasing function of time. This means that, in principle one can explain the observed luminosity with a small $W_{\rm sh}$ by taking longer dynamical time. However in practice, the dynamical time is not a free parameter, because it is well constrained by the bubble radius and shell speed. Therefore, the modeling of the ambient density profile is crucial to interpret $\gamma$-ray observation.

\subsubsection{Leptonic origin} \label{subsubsec:lepg}
Low energy photons ($\ll $ GeV) which come from stars and/or Cosmic Microwave Background (CMB) radiation can gain significant energy via inverse Compton scattering with relativistic electrons. These secondary photons can be a possible source of $\gamma$-rays in ISBs.

Suppose the incident photons are dominated by stellar radiation with energy $E_{\rm incident} \sim 0.01-100$ eV (far infrared to extreme UV). The corresponding Lorentz factor of relativistic electrons, require to enhance the energy of stellar photons to $E_{\rm obs}$ ($\approx 0.1-100$ GeV), is spread over $\Gamma\approx (E_{\rm obs}/E_{\rm incident})^{1/2}\sim 10^{3}( \Gamma_{\rm min}) -10^{6}(\Gamma_{\rm max})$. Assuming the number density distribution of relativistic electrons is $n(\Gamma)= \kappa_{1} \Gamma^{-p}$ ($p\approx 2.2$ is the spectral index of relativistic electrons), we estimate the $\gamma$-ray luminosity ($L^{\rm IC}_{\rm \gamma}$) from (see Equation $7.21$ in \citealt{Rybicki1979})
\begin{eqnarray} \label{eq:Lic1}
L^{\rm IC}_{\rm \gamma} & = &\int_{\rm V} dV\ \left[ \frac{4}{3}\,\sigma_{\rm T}\,c\,e_{\rm ph}\,\kappa_{\rm 1}\,\frac{\Gamma^{3-p}_{\rm max}-\Gamma^{3-p}_{\rm min}}{3-p}\right]\,
\end{eqnarray} 
where $e_{\rm ph}$ is the stellar radiation energy density and $\sigma_{\rm T} $ is the Thomson cross-section. The normalization constant $\kappa_{\rm 1}$ is obtained from the energy density of CR electron $e_{\rm cr\_e}$ as,
\begin{eqnarray} \label{eq:kappa1}
\kappa_{\rm 1} \approx \frac{e_{\rm cr\_e}}{m_{\rm e}c^{2}}(p-2)\left[\frac{1}{\Gamma^{p-2}_{\rm L}}-\frac{1}{\Gamma^{p-2}_{\rm U}}\right]^{-1}.
\end{eqnarray}
Here, the lower and upper cutoff of Lorentz factor can be set to $\Gamma_{\rm L}\rightarrow 1$ and $\Gamma_{\rm U }\rightarrow \infty$. We assume the energy density of relativistic electrons $e_{\rm cr\_e}=e_{\rm cr}(m_{\rm e}/m_{\rm p})^{(3-p)/2}$ \citep{Persic2014}. For $p\approx 2.2$, this gives $e_{\rm cr\_e}\approx 0.05\,e_{\rm cr}$. 

The stellar radiation energy density ($e_{\rm ph}$) depends on the distance from stars and radiation luminosity ($L_{\rm rad}$). Assuming that the stars are confined in a small region and that the total radiation luminosity $L_{\rm rad}\sim 500 L_{\rm w}$ ($L_{\rm w}$ is the wind power) [\citealt{Leitherer99}], $e_{\rm ph}$ at a distance $r$ can be obtained from,
\begin{eqnarray}
e_{\rm ph} (r) &=& \frac{L_{\rm rad}}{4\pi r^2 c} \nonumber \\ 
&\approx & 435 \left(\frac{L_{\rm w}}{5\times 10^{38}{\rm erg\,s^{-1}}}\right) \left(\frac{r}{10{\rm pc}}\right)^{-2}{\rm eV\,cm^{-3}},
\end{eqnarray}
which is much larger than the energy density in CMB photons $\sim 0.3\,{\rm eV\,cm^{-3}}$. Using Equation (\ref{eq:Lic1}), we find that the $\gamma$-ray luminosity in $0.1-100$ GeV energy due to inverse Compton scattering is
\begin{eqnarray} \label{eq:Lgic}
L^{\rm IC}_{\rm \gamma} & \approx & 172 \times 10^{-16}\left(\frac{L_{\rm w}}{5\times 10^{38}{\rm erg\,s^{-1}}}\right) \nonumber \\ & \times &\left[\int_{V} dV\, \left(\frac{r}{10{\rm pc}}\right)^{-2}\,e_{\rm cr}\right] {\rm erg\,s^{-1}}\, ,
\end{eqnarray} 
where $dV$ and $e_{\rm cr}$ are in CGS units. 

\begin{table*}
\begin{footnotesize}
\begin{center}
\caption{The output from star cluster observations} \label{tab:obs_data}
\begin{tabular}{l c c c c c c c c c c c }
  \hline\hline
  [1] & \multicolumn{2}{c}{$[2]$ Central source} &\multicolumn{2}{l}{$[3]$ Bubble}  & \multicolumn{2}{l}{$[4]$ $\gamma$-ray} & \multicolumn{2}{l}{$[5]$ Thermal X-ray} & \multicolumn{2}{l}{$[6]$ Radio} & $[7]$\\
\cline{2-3} \cline{4-5} \cline{6-7} \cline{8-9}  \cline{10-11}
Object  & $M_{\rm *}$ &  $L_{\rm w}$ & Age  & $R$ & $E_{\rm \gamma}$ &  $L_{\rm \gamma}$  &  $T_{\rm x}$ &  $L_{\rm x}$ & $\nu$ & $F_{\rm R}$ & Ref. \\
  name   &   \scriptsize{($M_{\rm \odot}$)} & \scriptsize{(erg s$^{-1}$)} & \scriptsize{(Myr)} & \scriptsize{(pc)} & \scriptsize{(GeV)}  & \scriptsize{(erg s$^{-1}$)} & \scriptsize{(${10^6}$K)} &  \scriptsize{(erg s$^{-1}$)} & \scriptsize{(GHz)} & \scriptsize{(Jy)}  &\\
    \hline
    \multirow{1}{*}{30Doradus} & \multirow{1}{*}{$5\times10^{5}$} & \multirow{1}{*}{$2\times10^{39}$}  & \multirow{1}{*}{$2$-$3$} & \multirow{1}{*}{$75$$-$$100$} & $0.1$-$20$ &  $\approx 1.4\times 10^{37}$ &  $4.5$ & $[4$$-$$7]\times10^{36}$ & $1.4$ & $56$ & a, b, c, d, e\\ 
    \multirow{1}{*}{Cygnus}  & $3\times 10^{4}$& $3\times10^{38}$ &$3$-$5$ & $\approx50$ & $1$$-$$100$ &  $[9$ $\pm$ $2]\times 10^{34}$ & -- & $[5$$-$$10]\times10^{35}$&-- & -- & f, g\\  
  \multirow{1}{*}{NGC 3603} & $\sim 10^{4}$ & $6\times10^{38}$ &$1$-$3$ & $\approx30$ & $1$-$250\, $ &  $\approx10^{36}$ &$6.2$ & $[2$$-$$5]\times10^{35}$& -- & -- & h, i, j, k\\ 
   \multirow{1}{*}{Westerlund1}  & $5\times 10^{4}$ & $\sim 10^{39}$ & $3$-$4$ & -- & $3$-$300$ &  $1.5\times10^{34}$& 6 &  $\lesssim 10^{34}$ & -- & -- & l, m\\ 
  \hline
\end{tabular}
\raggedright{\\
References: a. \citet{Abdo2010}, b. \citealt{Abramowski2015}, c. \citealt{Hughes2004}, d. \citealt{Knodlseder2013}, e. \citealt{Lopez2014}, f. \citet{Ackermann2011}, g. \citet{Wright2010}, h. \citet{Crowther1998}, i. \citet{Rosen2014},  j. \citet{Yang2017},\\ k. \citet{Harayama2014}, l. \citet{Muno2006}, m. \citet{Ohm2013}.}
\end{center}
\end{footnotesize}
\end{table*}
Taking\footnote{Observations of ISBs suggest that the column density is $L\sim 10^{21-22}\, {\rm cm^{-2}}$ (e.g. \citealt{Kim2003}; \citealt{Murphy2012}). For a typical ISB with radius, say $R \sim 10$ pc, number density $\approx L/R\sim 32-320 \,{\rm cm^{-3}}$.} $n_{\rm N}\approx 4 \rho_{\rm c}(r_{c}/r)^{s}/m_{\rm H}$ where $\rho_{\rm c}=220\,m{\rm _H\,cm^{-3}}$, $r_{\rm c} =5\,{\rm pc}$ and $s=1$ (c.f. Figure \ref{fig:ambprof}), Equations (\ref{eq:Lum_gamma}) and (\ref{eq:Lgic}) give the ratio of hadronic to Leptonic $\gamma$-ray luminosity:
\begin{equation} \label{eq:ghgl}
\frac{L^{\rm H}_{\rm \gamma}}{L^{\rm IC}_{\rm \gamma}}\approx 2.6 \left(\frac{\rho_{\rm c}}{220m{\rm _H\,cm^{-3}}}\right)\left(\frac{L_{\rm w}}{5\times 10^{38}{\rm erg\,s^{-1}}}\right)^{-1}\left(\frac{r}{10{\rm pc}}\right)\,.
\end{equation}
This suggests that both hadronic and leptonic interaction can be important to explain observed $\gamma$-ray photons in ISBs, although $L^{\rm H}_{\rm \gamma}$ dominates for large bubbles. 

\subsection{X-ray} \label{subsec: X-ray}
X-ray emissions depend on the inner structure of the ISB. For a qualitative understanding of X-ray luminosity ($L_{\rm x} $), we consider the emission to be due to thermal bremsstrahlung which yields,
\begin{eqnarray}\label{eq:Lx_c}
L_{\rm x} = \int_{V}dV \left[1.4\times 10^{-27}Z^{2}g_{\rm B}\,n_{\rm e}n_{\rm i}\,T^{1/2}\right] 
\end{eqnarray}
We take $Z\approx 1$, $g_{\rm B}=1.2$  and $n_{\rm e}\approx n_{\rm i}=P_{\rm th}/(k_{\rm B} T)$ and obtain
\begin{eqnarray} \label{eq:Lx_ap}
L_{\rm x} & \approx  &   3.7\times 10^{5}\,R^{3}\,T^{-3/2}\,P_{\rm th}^2\nonumber\\
&  \sim & 3.1\times 10^{34}\left(\frac{R}{10\rm pc}\right)^{3}\left(\frac{T}{5\times 10^{7}{\rm K}}\right)^{-3/2}\nonumber\\
& & \times \left(\frac{P_{\rm th}}{10^{-9}\rm cgs}\right)^{2}{\rm erg\,s^{-1}}
\end{eqnarray}
In case of CR acceleration, $P_{\rm th}$ will be smaller than in the one-fluid case, which may change $L_{\rm x}$. Therefore, the X-ray luminosity is an important diagnostic to identify a CR dominated bubble.
\subsection{Radio} \label{subsec:radiotheo}
We also wish to estimate the synchrotron emission rate from relativistic electrons. We consider the number density distribution of relativistic electrons to be $n(E)=\kappa_{\rm 2}\, E^{-p}$. Note that the normalization constant, $\kappa_{\rm 2}$ is different from $\kappa_{\rm 1}$ of Equation (\ref{eq:kappa1}). Denoting the magnetic field by $B$, the synchrotron volume emissivity is given by (see equation ($8.131$) in \citealt{Longair2011}),
\begin{eqnarray} \label{eq:synchro}
j_{\rm \nu} & \simeq & 2.3\times10^{-25}\ a(p)\, B^{(p+1)/2}\, \kappa^{\prime}_{\rm 2}\, \\ \nonumber
 &   & \times \left(\frac{3.217\times10^{17}}{\nu}\right)^{(p-1)/2}\ \, {\rm J\,s^{-1}\,m^{-3} Hz^{-1}}
\end{eqnarray}
Here $a(p)\simeq 0.45$ for $p=2.2$ (table $8.2$ in \citealt{Longair2011}), the magnetic field $B$ in Tesla, $\kappa_{\rm 2}\approx$ $\left[(p-2) \left(m_{\rm e}c^{2}\right)^{p-2}\left(e_{\rm cr\_e}\right)\right]$ in ${\rm J^{p-1} m^{-3}}$, and $\kappa^{\prime}_{\rm 2} $ is obtained from $ \kappa_{\rm 2}$ after a unit conversion to ${\rm (GeV)^{p-1} m^{-3}}$.

Therefore, the luminosity per unit frequency is
\begin{eqnarray}  \label{eq:synchro_approx}
\frac{d L_{\rm R}}{d\nu} & = & \int_{\rm v}\, dV j_{\rm \nu} \\ \nonumber
& \sim & 1.4\times 10^{24}\left(\frac{R}{10\,{\rm pc}}\right)^{3}\left(\frac{B}{{40\rm \mu G}}\right)^{1.6}\\ \nonumber
 &   & \times \left(\frac{e_{\rm cr\_e}}{10^{-10}{\rm cgs}}\right)\left(\frac{\nu}{\rm 1.4 GHz}\right)^{-0.6}\ {\rm erg\,s^{-1}\,Hz^{-1}}
\end{eqnarray}

In the following sections we use numerical simulations to determine these observables using more realistic analysis.
\begin{figure*}
\centering
\includegraphics[height= 2.2in,width=6.9in]{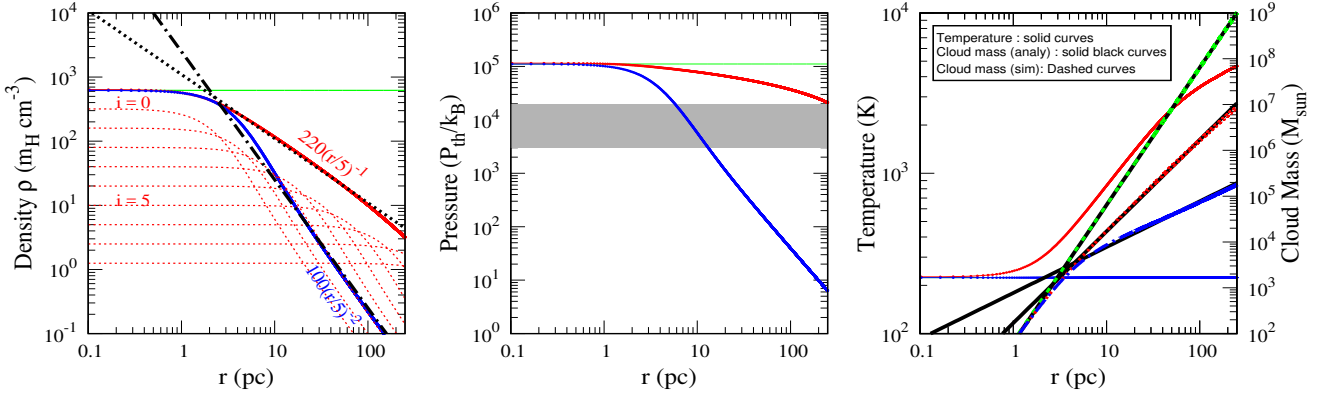}
\caption{Comparison of three different cloud profiles. Green and blue curves denote a uniform and a non-singular self-gravitating isothermal ambient medium respectively. Red curves represent the ambient medium used in this work. The grey shaded region in the middle panel shows the average thermal pressure ($P\sim G \Sigma^{2}$) observed in molecular clouds (\citealt{Hughes2010}). In the right-most panel, dashed (solid black) curves show the cloud mass for respective profiles obtained numerically (analytically) i.e., for $R_{\rm cl}=250$ pc, $M_{\rm cl}\simeq 10^{9}$, $9\times 10^{6}$ and $1.7 \times 10^{5}$ M$_{\rm \odot}$ respectively.}
\label{fig:ambprof}
\end{figure*}
%
\section{Observations of ISBs} \label{sec:obs}
In Table \ref{tab:obs_data}, we show the results from multi-wavelength observation of four massive star clusters. Column [2] shows that the wind power ranges between $10^{38}\lesssim L_{\rm w}/({\rm erg\,s^{-1}})\lesssim 10^{39}$. Column [3] shows the radius ($R$) of the bubble ($\sim 10-100$ pc) and their dynamical age ($\lesssim 5$ Myr). The details of $\gamma$-ray and X-ray observations are listed in columns [4] and [5] respectively. These indicate that $\gamma$-ray luminosity $(L_{\rm \gamma})\lesssim 10^{-2} L_{\rm w}$ and the X-ray luminosity $L_{\rm X}/L_{\rm w} \sim 10^{-3} - 10^{-2}$. For all sources, the $\gamma$-ray spectral index in $0.1-200$ GeV energy band is $\approx 2.2$. Column [6] shows that the radio power from 30 Doradus at $1.4$ GHz is $dL_{\rm R}/d\nu=4\pi D^{2} F_{\rm R}\sim 1.7\times 10^{26}\,{\rm erg\,s^{-1}\, Hz^{-1}}$ (by taking $D\approx 50$ kpc) [\citealt{Hughes2004}; see also Figure $5$ in \citealt{Foreman2015}].

Note that, out of these objects, $30$ Doradus is the only one in which most of the massive stars are located at the center and the structure of the bubble is close to spherical. This motivates us to compare our results with $30$ Doradus, which is discussed in \S \ref{subsec:compobs}.
\section{Simulation set-up} \label{sec:sim_setup}
We use a modified version of the PLUTO to perform hydrodynamic simulations in the presence of a CR fluid (\citealt{Mignone2007}; Gupta et al, in preparation). The following equations are solved:
 \begin{eqnarray} \label{eq:mass}
\frac{\partial \rho}{\partial t}+\vec{\nabla}.(\rho\,\vec{v}) & = & S_{\rm \rho} \\
\label{eq:momentum}
 \frac{\partial }{\partial t} (\rho\,\vec{v} ) +\vec{\nabla}.(\rho\,\vec{v}\otimes\vec{v}+p_{\rm tot})  & = & \rho \vec{g} \\
 \label{eq:totenergy}
 \frac{\partial e_{\rm tot}}{\partial t}+\vec{\nabla}.\left[ \left(e_{\rm tot} + p_{\rm tot}\right)\vec{v} + \vec{F}_{\rm t}+\vec{F}_{\rm crd} \right] & = & \rho\vec{v}.\vec{g}  \\ \nonumber & &  + S_{\rm e} -q^{\rm eff}_{\rm th}\\
 \label{eq:energycr}
 \frac{\partial e_{\rm cr}}{\partial t}  +\vec{\nabla}.\left[ \left(e_{\rm cr}+ p_{\rm cr}\right)\,\vec{v}+\vec{F}_{\rm crd}\right]  & = &\vec{v}. \vec{\nabla} p_{\rm cr} \\ \nonumber & & + S_{\rm cr} - q_{\rm cr}
\end{eqnarray}
Here $\rho$ and $\vec{v}$ are the mass density and fluid velocity respectively, $p_{\rm tot}=p_{\rm th}+p_{\rm cr}$ is the sum of thermal and CR pressures, $e_{\rm tot}$ is the sum of kinetic ($e_{\rm k}$), thermal ($e_{\rm th}$) and CR ($e_{\rm cr}$) energy densities. The adiabatic index for the respective fluids are chosen as $\gamma_{\rm th,cr}= 5/3,4/3$. We have used HLL Riemann solver, piecewise linear reconstruction and RK$2$ time stepping. The CFL number is taken as $0.3$.
\subsection{Ambient medium} \label{subsec:ambient}
The typical size of giant molecular cloud is $\sim 10-100$ pc and masses are $\sim 10^{4}-10^6$ M$_{\rm \odot}$. Detailed observations suggest that the cloud mass and radius follow $M_{\rm cl}\propto R_{\rm cl}^{2}$, i.e., the density profile ($\rho$) $\propto  r^{-1}$ (\citealt{Solomon1987}; \citealt{Hughes2010}; \citealt{Pfalzner2015}). In order to model this, we consider a self-gravitating gas cloud.

The most popular choice for a self-gravitating cloud is an isothermal sphere. A fit for the density profile in this case is given by \citet{Natarajan1997},
\begin{equation}
\rho(r,r_{\rm c}) =\rho_{c}\left[ \frac{5}{1+(r/r_{\rm c})^{2}/10}-\frac{4}{1+(r/r_{\rm c})^{2}/12}\right]\,
\end{equation}
Here $r_{\rm c}=c_{\rm s}/(4\pi G \rho_{\rm c})^{1/2}=\left[k_{\rm B}T/(4\pi G \rho_{\rm c}\mu m_{\rm_H})\right]^{1/2}\simeq 2.2\, T^{1/2}_{\rm 2}\,\rho^{-1/2}_{c,\rm 2}$ pc is the core radius, $T$ is the temperature, $\rho_{\rm c}$ is the core density and $\mu=1.26$ (cold neutral medium). However, this profile does not give $\rho \propto r^{-1}$. We, therefore, relax the isothermal assumption on the global length-scale ($\sim 100$ pc) of the cloud. Instead, we add several self-gravitating isothermal clouds and obtain a resultant density profile from,
\begin{equation}
\rho(r) = \sum\limits_{i=1}^{n} \rho(r,r^{\rm i}_{\rm c})\,
\end{equation}
where we set the core density and temperature of the clouds as
\begin{eqnarray}
\rho_{\rm c}^{\rm i} = 2^{5-i}\ 10\,m{\rm _H\, cm^{-3}}\ , \ T^{\rm i} = \frac{1600}{2^{5-i}}\ {\rm K}
\end{eqnarray}
where $i=0,1,2,...,8$ ($n=8$). This profile provides a dense core ($\approx 620\, m{\rm _H\,cm^{-3}}$) with temperature $\approx 200$ K and a mean surface density $\Sigma\approx 50\, M{\rm _\odot\,pc^{-2}}$, see the comparisons of different ambient profiles in Figure \ref{fig:ambprof}. 

To maintain hydrostatic equilibrium, we take into account the self gravity of the individual clouds. The net gravitational acceleration $\vec{g}$ (see Equations (\ref{eq:momentum}) and (\ref{eq:totenergy})) is obtained as,
\begin{eqnarray}
\vec{g}(r) = \sum \limits_{i=1}^{n} \frac{\rho(r,r^{\rm i}_{\rm c})}{\rho(r)}\left[\frac{(c^{\rm i}_{\rm s})^2}{\rho(r,r^{\rm i}_{\rm c})}\frac{d}{dr}\rho(r,r^{\rm i}_{\rm c})\right] \hat{r}
\end{eqnarray}
We find that the ambient profiles are stable for a few hundred Myr.

Note that the cloud profile obtained here is not unique. One can choose a different set of parameters to obtain different ambient density profiles. Furthermore, in a realistic scenario, the ambient medium consists of high density clumps ($\gtrsim 10^{4}\, m{\rm_{H}\,cm^{-3}}$). Therefore, our ambient profile should be treated as a directionally averaged cloud profile.
\subsection{Wind-driving region} \label{subsec:wind}
For the runs performed in 1D spherical geometry, we choose a spherical region of radius $r_{\rm inj}=1$ pc around $r=0$ and set a fine spatial resolution ($\Delta r = 0.05$ pc). This allows us to minimize nonphysical cooling losses at the early stages of shock formation (see section $4$ in \citealt{Sharma2014}, also see Equation ($10$) in \citealt{Gupta2016}). In our fiducial set-up, we set $\dot{M}=4\times 10^{-4}\,M{\rm _\odot\,yr^{-1}}$ and $L_{\rm w}=5\times 10^{38}\,{\rm erg\,s^{-1}}$ which have been added uniformly (i.e, $S_{\rm \rho}=\dot{M}/V_{\rm inj}$ and $S_{\rm e}=L_{\rm w}/V_{\rm inj}$ where $V_{\rm inj}=4\pi r_{\rm inj}^3/3$). Therefore, at the sonic point ($r=1$ pc), the wind velocity is $1414\,{\rm km\,s^{-1}}$ which asymptotically approaches $v_{\rm w}=(2L_{\rm w}/\dot{M})^{1/2}\approx 2000\,{\rm km\,s^{-1}}$ (\citealt{CC1985}). We discuss the dependence of our results on these parameters in section \ref{subsec:vw}.

To test the reliability of our fiducial 1D model, we perform 3D simulation, particularly to study the effects of distributed stars. For these runs we use Cartesian geometry and distribute a total $N_{\rm *}=500$ (assumed) injection points by using a Gaussian random number generator with zero mean value and the standard deviation of $1$ pc (c.f. Figure \ref{fig:disstars}). The radius of the injection points is taken as $\delta r_{\rm inj}=0.3$ pc, where mass and energy are added uniformly (similar to $1$D). The spatial resolution in the central region, [$(x,y,z)\in (-5,5)$ pc] which covers all injection points, is set to $0.125$ pc.
\subsection{CR injection} \label{subsec:crinject}
We use the following two scenarios for CR injection:
\begin{itemize}
\item Injection in the wind-driving (IWD) region: Internal shocks in the wind-driving region can be efficient site for CR acceleration. However, it is difficult to spatially resolve them. To investigate this type of acceleration scenario, we use a parameter $\epsilon_{\rm cr}$ to denote the fraction of wind energy injected into CRs. The fiducial value is $\epsilon_{\rm cr}=0.1$.

\item Injection at the shock (ISH): In this case, we have injected CRs directly at the resolved shocks (i.e. at forward and reverse shock of the ISB). To identify whether a computation zone is shocked or not, we use the following conditions.\\
\\
(i) $\vec{\nabla}.\vec{v} < 0$,\\
\\
(iii) $\Delta x |\vec{\nabla}p|/p\geq\delta_{\rm tolerance}$\\
\\
(iii) $\vec{\nabla}T.\vec{\nabla}\rho > 0$.

In this work we have taken $\delta_{\rm tolerance}=1.5$. The last condition helps to exclude spurious oscillations at the contact discontinuity which can be detected as a shock (\citealt{Pfrommer2017}). We then find the total non-kinetic energy density of the shocked zone (i.e., $e_{\rm th}+e_{\rm cr}$) and re-distribute it by a parameter $\epsilon^{\rm ISH}_{\rm cr}$ such that the CR pressure fraction of the shocked zone $w= p_{\rm cr}/(p_{\rm th}+p_{\rm cr})=\epsilon^{\rm ISH}_{\rm cr}/(2-\epsilon^{\rm ISH}_{\rm cr})$.

Note that the fraction of energy transfer depends on the location of the grid point, which is not necessarily the peak location (density/pressure) of a shock. This may reduce the effective post shock CR pressure (which determines the CR pressure fraction $W_{\rm in}$/$W_{\rm sh}$ in the interior/shell) from the injected value ($w$).

\end{itemize} 
In both injection models, we ensure that CR injection does not add any additional energy in the computational zone. We simply distribute a fraction of the mechanical energy (by using $\epsilon_{\rm cr}$ or $w$) in the form of CRs either in the wind-driving region or at the shocks.
\begin{table}
\caption{Simulation parameters.}
\begin{center}
\begin{tabular}{l c c}
  \hline\hline
  Parameter & Fiducial & Range covered\\ 
  \hline
  $L_{\rm w}\,({\rm erg\,s^{-1}})$ & $5\times 10^{38}$ &  $10^{38}-10^{39}$\\
  $v_{\rm w}\,({\rm km\,s^{-1}})$ & $2000$ &  $\approx 1000-5000$\\
  \hline
  $\epsilon_{\rm cr}$ & $0.1$ &  $0.01-0.20$\\
  $w$ & $0.33$ &  $0.05-0.54$\\
  $\kappa_{\rm cr}\,({\rm cm^2\,s^{-1}})$ & $5\times 10^{26}$ &  $5\times 10^{25}- 3\times 10^{27}$\\
  \hline
  Resolution in $1$D (pc) & $0.05^{*},0.06$ & $0.03-0.50$\\
  Resolution in $3$D (pc) & $0.125^{*},0.79$ & $-$\\
    \hline
\end{tabular}\\
\raggedright{$^{*}$ Resolution in the central region (\S \ref{subsec:wind}).}
\end{center}
\label{tab:rundetails}
\end{table}
\begin{figure*}
\centering
\includegraphics[height= 2.8in,width=7.2in]{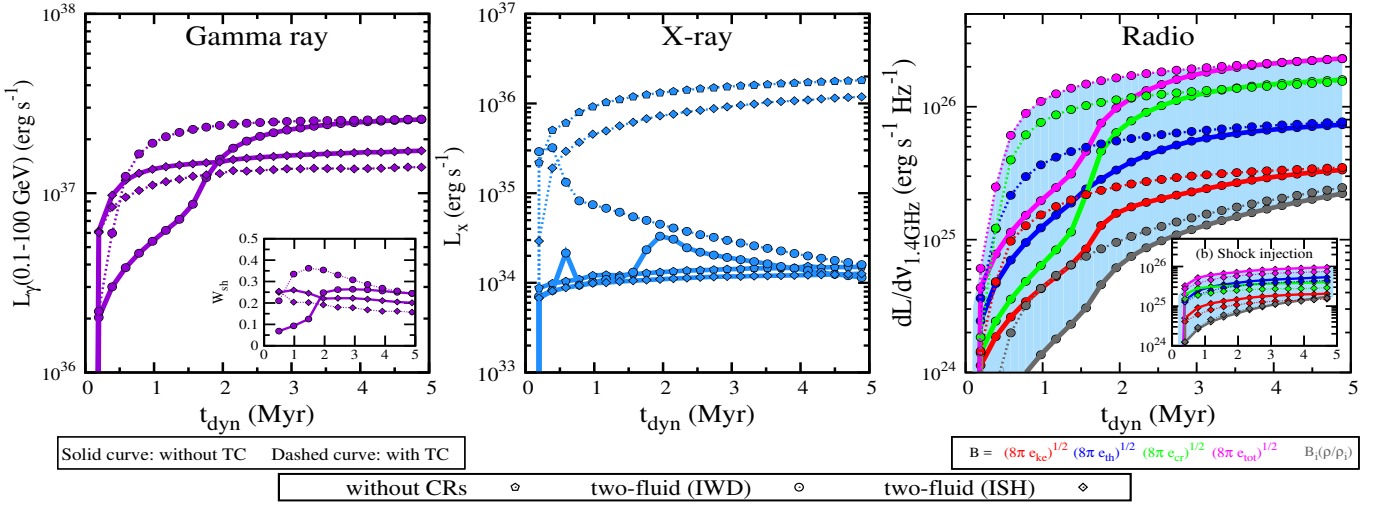}
\caption{Time evolution of $\gamma$-ray (violet), X-ray (blue) and Radio luminosities for our fiducial runs (see Table \ref{tab:rundetails}). Three different point styles, pentagon and diamond/circle, are used to indicate one-fluid and two-fluid (model: IWD/ISH) ISB respectively. Dashed/solid line represents runs with/without thermal conduction. The sky-blue shaded region in the right-most panel displays the expected radio emission at $1.4$ GHz. The figure indicates that luminosities after $\gtrsim 2.5$ Myr do not change significantly.}
\label{fig:lum}
\end{figure*}

\subsection{Microphysics} \label{subsec:micro}
\subsubsection{Cooling losses and heating}
Cooling loss of the thermal fluid is taken into account by using a tabulated cooling function for the gas metallicity $Z=0.4\,Z_{\rm \odot}$. To mimic photo-ionization heating from the central radiation field, we turn off cooling when temperature $T<10^{4}$ K.

The cooling loss rate of CR fluid due to the hadronic and Coulomb interactions is taken to be $q_{\rm cr}=7.5\times 10^{-16}\,n_{\rm H}\,e_{\rm cr}\, {\rm erg\,cm^{-3}\,s^{-1}}$ (see section 2.1 in \citealt{Guo2008}). The corresponding collisional heating rate of thermal gas is given as $2.6\times 10^{-16}\,n_{\rm H}\,e_{\rm cr}\, {\rm erg\,cm^{-3}\,s^{-1}}$. Therefore, $q^{\rm eff}_{\rm th}$ and $q_{\rm cr}$ in Equations (\ref{eq:totenergy}) and (\ref{eq:energycr}) are
\begin{eqnarray}
q^{\rm eff}_{\rm th} & = & \Lambda_{\rm N} n_{\rm i} n_{\rm e} - 2.6\times 10^{-16}\,n_{\rm e}\,e_{\rm cr}\, {\rm erg\,cm^{-3}\,s^{-1}}\\ 
q_{\rm cr} & = & 7.5\times 10^{-16}\,n_{\rm e}\,e_{\rm cr}\, {\rm erg\,cm^{-3}\,s^{-1}}
\end{eqnarray}
Note that heating due to CR steaming may affect the thermal fluid more than collisional heating. However, it is not possible to include it in our hydrodynamic set-up. Further, we find that the effect of CR collisional heating is negligible in our set-up. The CR heating can be better studied with the help of MHD simulations.

\subsubsection{Thermal conduction \& CR diffusion}
We assume that both thermal conduction (hereafter, TC) and CR diffusion are isotropic. We use thermal conduction (TC) to have the Spitzer value ($6\times10^{-7}\,T^{5/2}$ in CGS) and also assume the saturated thermal conduction (see section $4.3$ in \citealt{Gupta2016}). The fiducial value of CR diffusion coefficient is set to $\kappa_{\rm cr}=5\times10^{26}$ cm$^{2}$ s$^{-1}$, unless otherwise mentioned (\citealt{Gupta2018}). For both cases, we choose STS method (\citealt{Alexiades1996}) to speed up the diffusion module.

\section{Results} \label{sec:result}
In this section, we show the results from fiducial runs (see Table \ref{tab:rundetails}). We first discuss $1$D simulations (section \ref{subsec:1d}) and then compare them with $3$D simulations (section \ref{subsec:3d}).
\subsection{1D runs} \label{subsec:1d}
The structural difference between one-fluid and two-fluid ISBs has been discussed in \citet{Gupta2018} (see their section $4.2$). Here we present the time evolution of multi-wavelength luminosities.
\subsubsection{$\gamma$-ray luminosity} \label{subsec:gamma_sim}
To obtain the $\gamma$-ray luminosity, we use Equation (\ref{eq:Lum_gamma}) and display the results in the left-most panel of Figure \ref{fig:lum}. The dashed and solid curves represent model with and without thermal conduction (TC).

The solid curve (without TC) in this figure shows that the $\gamma$-ray luminosity (hereafter, displayed by violet curves) in both injection models (denoted by circular symbol: IWD and diamond symbol: ISH) is an increasing function of time. This is expected because, as time evolves, the swept-up mass (in the shell) increases. We also see that the $\gamma$-ray luminosity for IWD model (circular symbols) evolves differently from ISH model (diamond symbols). This can be understood from Equation (\ref{eq:Lum_g_f}) which shows
$L_{\rm \gamma}\propto W_{\rm sh}\,t_{\rm dyn}^{1/4}$. The parameter $W_{\rm sh}$ is the source of difference between the IWD and ISH models for the following reasons.

For IWD, when the Mach number of reverse shock becomes $\gtrsim 12$, most of the upstream kinetic energy is converted into CRs (\citealt{Becker2001}; also see section $4.2$ in \citealt{Gupta2018}). When TC is off (solid curve), this results in a large increase in CR pressure downstream of the reverse shock after $t_{\rm dyn}\gtrsim 2$ Myr. These CRs diffuse and increase the CR pressure in the shell. Therefore, in the early stages of evolution, $W_{\rm sh}$ increases with time. This is illustrated in the subplot of the same panel. The run with TC (dashed curve) shows a similar result but with an earlier rise than without TC ($t_{\rm dyn}\gtrsim 0.5$ Myr). 
 
On the contrary, for the shock injection scenario (ISH), $W_{\rm sh}$ is fixed. This causes a slower change with time. In this case, $L_{\rm \gamma}$ hardly shows any difference between with and without TC (compare the diamond symbols connected by solid and dashed lines). 

\subsubsection{X-ray luminosity} \label{subsec:xray_sim}
We use the Mekal plasma model (for gas metallicity $Z=0.4\,Z_{\rm \odot}$) to estimate the X-ray luminosity in ($\approx0.5-2$) keV energy band and the results are displayed in the middle panel of Fig. \ref{fig:lum}.

Without TC (solid curves), the X-ray luminosity for all models is $\sim 10^{34} \,{\rm erg\,s^{-1}}$ ($\sim 2\times 10^{-5}\,L_{\rm w}$ ). To illustrate this, we recall Equation (\ref{eq:Lx_ap}) which yields,
\begin{eqnarray}\label{eq:Lxsim}
\frac{L_{\rm x}}{\rm erg\,s^{-1}} & \approx & 1.7\times 10^{37}\, T^{-3/2}_{\rm 7} \left[\left(\frac{\gamma-1}{9\gamma-5}\right)^{5/4}\, (1-W_{\rm in})^2\right]\nonumber \\
& & \times  L_{\rm 38}^{5/4}\, (\rho_{\rm c,220}\,r_{\rm c,5})^{3/4}\,t_{6}^{-1/4}
\end{eqnarray} 
where $T$ and $W_{\rm in}$ denote the volume averaged temperature and CR pressure fraction inside the bubble respectively.

Without CRs, for our fiducial parameter $T_{\rm 7}\approx 5$, Equation (\ref{eq:Lxsim}) gives $L_{\rm x}\approx 2.9\times 10^{35}\, {\rm erg\,s^{-1}}$ at $t_{\rm dyn} \approx 3\,{\rm Myr}$. However, in the simulation we find $L_{\rm x}\simeq 1.5\times 10^{34}{\rm\,erg\,s^{-1}}$. The difference arises because $T$ is outside of our range of interest ($\approx 0.5-2$ keV) and also because of cooling losses in the bubble (c.f. Figure \ref{fig:bubble_size}). 
\begin{figure}
\centering
\includegraphics[height= 2.45in,width=3.35in]{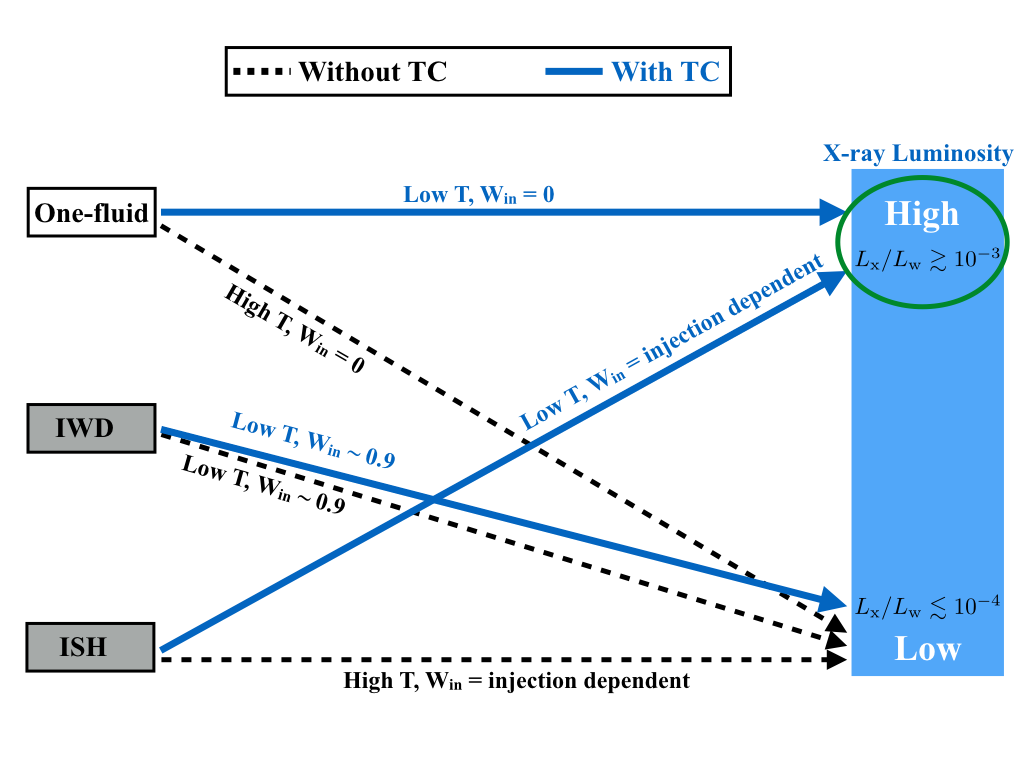}
\caption{Schematic diagram of $L_{\rm x}$ ($\propto T^{-3/2}(1-W_{\rm in})^2$) in different models. In IWD scenario we have only shown the case when the reverse shock is dominated by CRs due to globally smooth solution. In this case it is not possible to produce high $L_{\rm x}$. The green circle represents  $L_{\rm x}$ corresponding to observation.}
\label{fig:illus}
\end{figure}

The variation of $L_{\rm x}$ in different models is schematically shown in Figure \ref{fig:illus}. For the IWD model, due to an efficient energy transfer from thermal to CR fluid at the reverse shock, the temperature reduces to $T_{\rm 7} \sim 0.4$ and $(1-W_{\rm in})^2\sim 10^{-2}$ (see figures $9$ and $10$ in \citealt{Gupta2018}), leading to $L_{\rm x}\approx 2 \times 10^{34}\,{\rm erg\,s^{-1}}$. In contrast, $L_{\rm x}$ in ISH model depends on $W_{\rm in}$ (a larger $w$ corresponds to smaller $L_{\rm x}$) (see diamond and pentagon symbols in Figure \ref{fig:lum}). 

A noticeable difference between IWD and ISH models is found when we include TC, displayed by the dashed curves in the middle panel of Fig. \ref{fig:lum}. In the absence of CRs, TC reduces the temperature without affecting the thermal pressure of the SW region. This increases $L_{\rm x}$ (see Equation (\ref{eq:Lxsim}) with a smaller $T_{\rm 7}$). With CR in the IWD model, the X-ray luminosity is $\gtrsim 10$ times smaller than ISH and one-fluid models. This is because of diffusive acceleration at the reverse shock which diminishes the shocked wind temperature and also reduces the effect of thermal conduction. For ISH model, $L_{\rm x}$ depends on $W_{\rm in}$, and for our choice of $w=0.33$, $L_{\rm x}$ can be large ($L_{\rm x}\gtrsim 10^{-3} L_{\rm w}$).
\subsubsection{Radio} \label{subsec:radio}
We use Equation (\ref{eq:synchro}) to model the synchrotron radio emission. Since we do not include magnetic field ($B$) in our runs, we use two different methods to estimate the magnitude of $B$. 

The first method uses equipartition of magnetic energy with kinetic (ke)/thermal (th)/CR (cr)/total energy (tot). The second method is motivated by observations that suggest that the magnetic field in a cloud depends on density \citep{Valle1993}. Therefore, the magnetic field ($B$) has been estimated using,

\begin{equation}
|\vec{B}| \approx
\left\{
	\begin{array}{ll}
		\sqrt{8\pi e_{\rm x}}\ \, {\rm where\,x: ke/th/cr/tot} \\
		B_{\rm i} \left(\rho/\rho_{\rm i}\right) &
	\end{array}
\right.
\end{equation}
where, the subscript `${\rm i}$' stands for the initial ambient value. For simplicity, we assume $B_{\rm i} =10\,{\rm \mu G}$ to be uniform. Due to ambiguity in magnetic field, we have five degenerate values of $L_{\rm R}$ ($=\nu dL_{\rm \nu}/d\nu$), at a given time. The result is shown by different colours in the right-most panel of Figure \ref{fig:lum}. Here, we use the same symbols (line styles) to represent IWD and ISH (with/without TC) models. 

For IWD model, $L_{\rm 1.4GHz}$ is consistent with the analytical estimates (Equation (\ref{eq:synchro_approx})). The subplot (b) for ISH model shows that the results depend weakly on time. The difference between IWD and ISH models stems for the fact that the fractions $W_{\rm in}$ and $W_{\rm sh}$, which determine the CR electron energy density ($e_{\rm cr\_e}$), evolve differently in these two models. For details see sections \ref{subsec:gamma_sim} and \ref{subsec:xray_sim}.

The luminosities in all bands show a weak time dependence after $\approx 2.5$ Myr. This allows us to compare with observations and explore the parameter dependence without invoking a particular epoch (c.f. section \ref{sec:discussion}).

\begin{figure}
\centering
\includegraphics[height= 3.1in,width=3.45in]{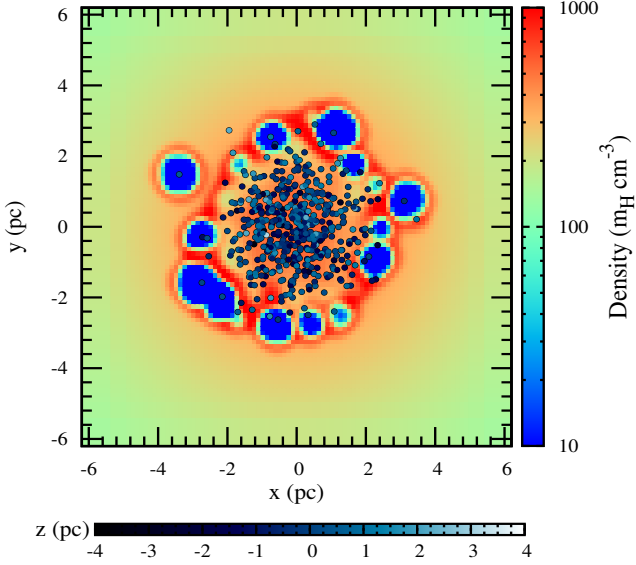}
\caption{Distributed injection in $3$D. The horizontal colour palette displays z-coordinate of the injection points (darker points are in the back, and brighter points in front). The vertical colour palette displays the snap shot of density profile in $z=0$ plane at $0.05$ Myr. The figure represents the early stage of bubble formation.}
\label{fig:disstars}
\end{figure}
\subsection{3D runs} \label{subsec:3d}
\subsubsection{Structure and dynamics}

\begin{figure}
\centering
\includegraphics[height= 6.4in,width=3.4in]{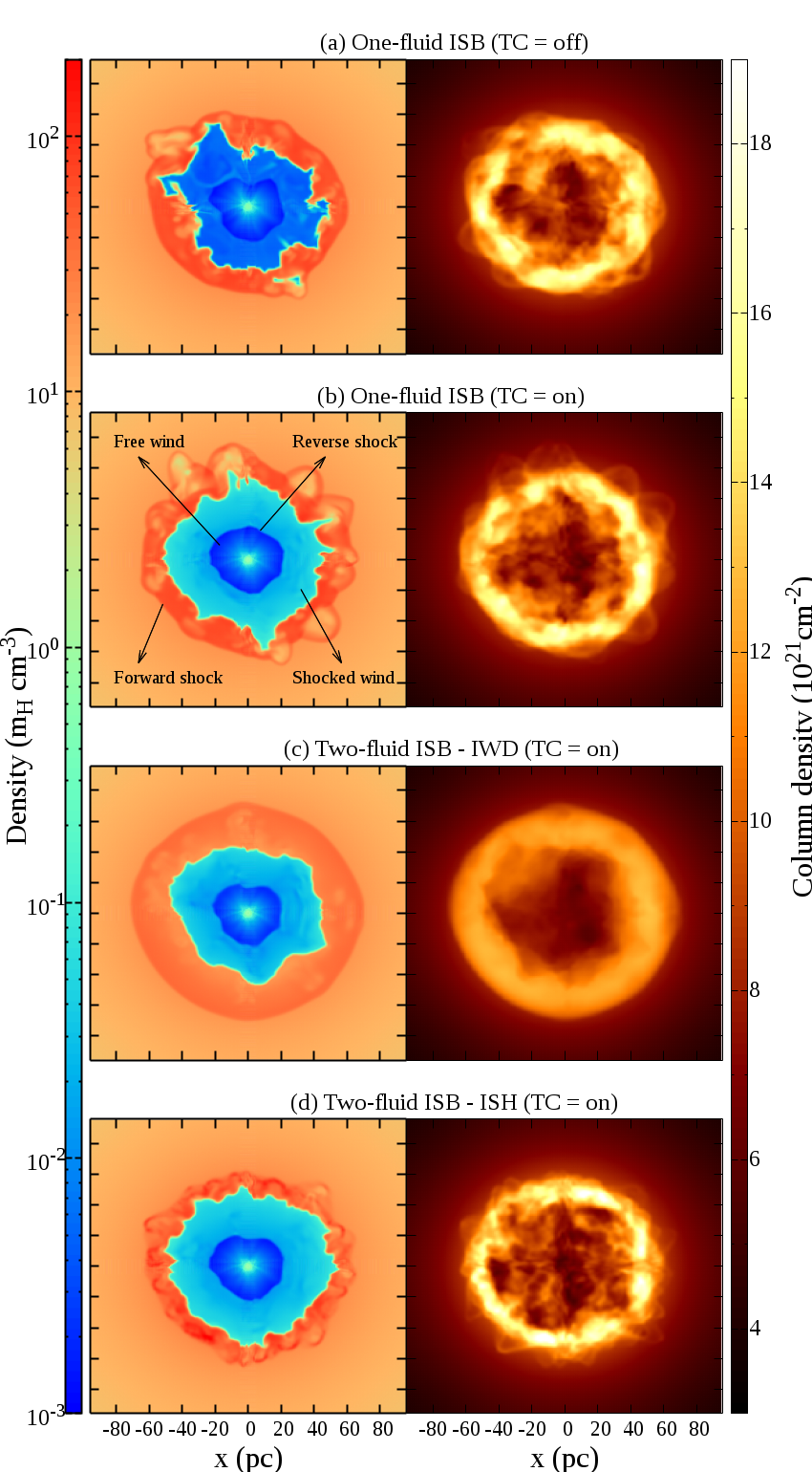}
\caption{The comparison of density profile at $z=0$ plane (left panel) and the column density along the $z$ axis (right panel) at $2.5$ Myr. The left panels clearly indicate the four distinct regions of the ISB. The right panels show that information of the internal structure is apparently lost due to projection effect.}
\label{fig:3dprofiles}
\end{figure}

\begin{figure}
\centering
\includegraphics[height=2.7in,width=3.4in]{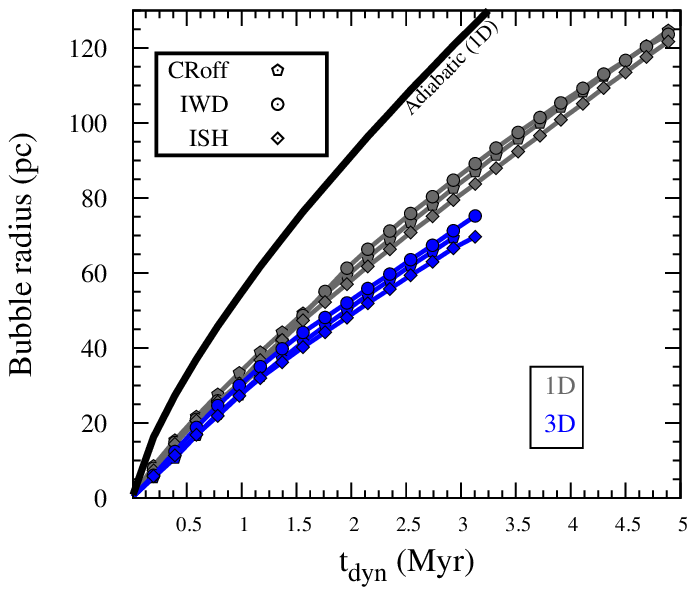}
\caption{Evolution of bubble radius in $1$D and $3$D. The comparisons of solid black curve (which stands for an one-fluid adiabatic run i.e. cooling and CRs both are not included) with other curves show that cooling losses have reduced the bubble size. Blue and grey curves (where micro-physics are included, \S \ref{subsec:micro}) show that the radius in $3$D is smaller than $1$D. The comparison of different symbols (having the same colour) indicates that the dynamical impact of CRs is negligible.}
\label{fig:bubble_size}
\end{figure}
To present a more realistic scenario, we perform $3$D simulations with the same fiducial parameters (see Table \ref{tab:rundetails}). In these runs, mass and energy are injected in a distributed manner. Figure \ref{fig:disstars} displays the injection points where the horizontal colour palette represents the $z$ coordinates of those points. The vertical colour palette displays the density snapshot in the $z=0$ plane at $0.05$ Myr. This shows that individual bubbles have started to merge at this epoch. At a later time ($t_{\rm dyn}\gtrsim 0.5$ Myr), the structure appears as an ISB, as shown in Figure \ref{fig:3dprofiles}.

The left panel of Figure \ref{fig:3dprofiles} shows the snapshot of density profile at $z=0$ plane for different models. The cumulative effect of all injection points produces a free wind profile followed by a reverse shock, shocked wind and forward shock. Therefore, the structure is very similar\footnote{It is worth noting that the coherent reverse shock may be destroyed if energy is injected via exploding supernovae rather than our smooth stellar winds (see \citealt{Sharma2014}; \citealt{Yadav2017}).} to that of a classical ISB.

The size of the bubble carries useful information, e.g. the mass of the shell, the volume of X-ray cavity. Because of the distributed nature of injection points in $3$D runs, the size evolution may be different. We show the comparison of $1$D (grey curves) and $3$D (blue curves) runs in Figure \ref{fig:bubble_size}. For both geometries (i.e. spherical and Cartesian), we first estimate the swept-up mass ($M_{\rm sh}$) and then we obtain the average shell radius by using:
\begin{equation}
R\approx\left[\frac{3-s}{4\pi\rho_{\rm c}r_{\rm c}^s} M_{\rm sh}\right]^{1/(3-s)}\ ,
\end{equation}
where $\rho_{\rm c}=220\,m{\rm _H\,cm^{-3}}$,  $r_{\rm c}=5\,{\rm pc}$ and $s=1$ (see Figure \ref{fig:ambprof}). From this figure we find that the radius of the bubble in $3$D runs is {\it smaller} compared to $1$D runs. Therefore, the $3$D runs are expected to show a {\it lower} luminosity. Otherwise, the different CR injection models do not show significant change from the one-fluid bubble. Therefore, the dynamical impact of CRs in ISBs may not be important.

\subsubsection{Time evolution of luminosities}
Following the methods described in sections \ref{subsec:gamma_sim}, \ref{subsec:xray_sim} and \ref{subsec:radio}, we estimate $L_{\rm \gamma}$, $L_{\rm x}$ and $L_{\rm R}$ for our $3$D runs. For $\gamma$-ray, we compute both hadronic and leptonic components.
\begin{figure}
\centering
\includegraphics[height= 5.8in,width=3.2in]{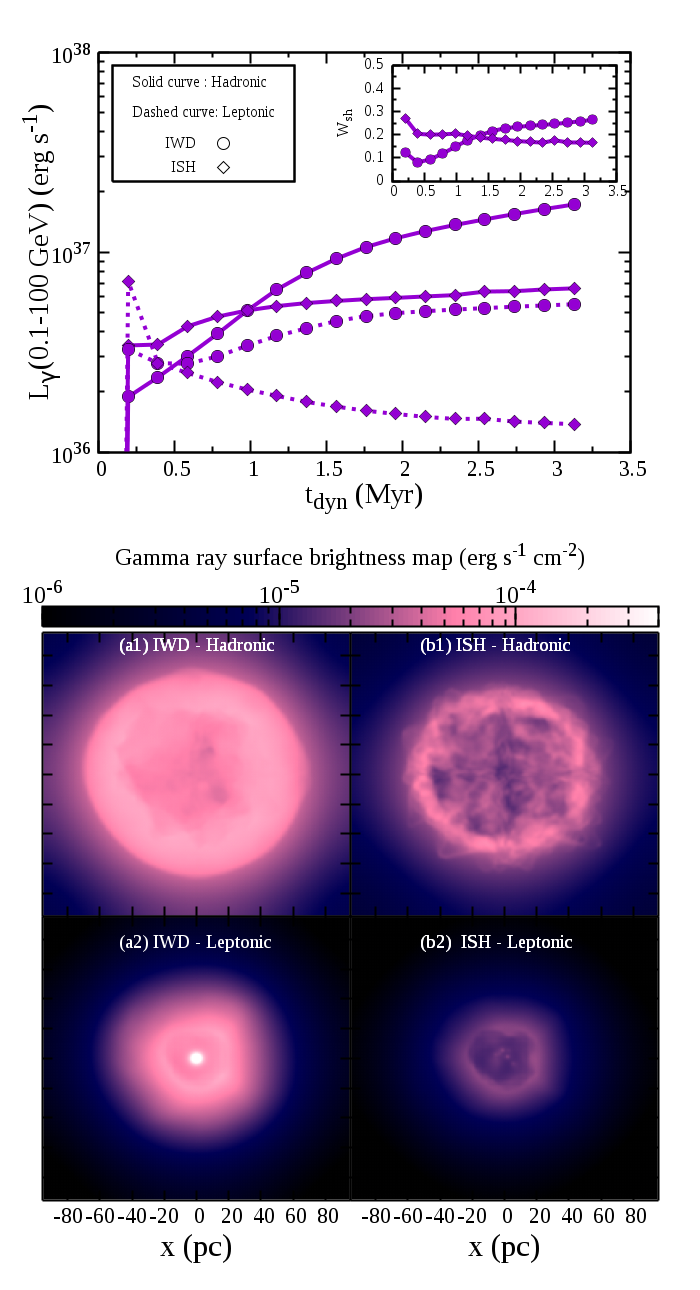}
\caption{[Top panel] Time evolution of $L_{\rm \gamma}$ in $3$D runs (with TC). The comparison of dashed curves with solid curves for a same symbol (circular: IWD and diamond: ISH) indicates that the inverse Compton scattering is subdominant compared to hadronic interaction (consistent with Equation \ref{eq:ghgl}). The subplot displays the CR pressure fraction in the shell.  [Bottom panel] The $\gamma$-ray surface brightness map along $z$ direction (Equation (\ref{eq:SBg})) at $2.5$ Myr.}
\label{fig:3dgamma}
\end{figure}

In Figure \ref{fig:3dgamma}, the solid curves displaying $L_{\rm \gamma}$ due to hadronic interaction show a similar time dependence as in $1$D. The CR pressure fraction (see subplot) for the models IWD and ISH are $W_{\rm sh}\approx0.25$ and $W_{\rm sh}\approx 0.17$ respectively. The luminosity is somewhat lower than the $1$D cases. This is because the size of the ISB is smaller than in $1$D model\footnote{Moreover, due to a smaller box size ($2 |L|$, spanning from $-100$ to $100$ pc), the ambient contribution is not completely captured in our analysis} (Figure \ref{fig:bubble_size}). The dashed curves show that inverse Compton scattering is sub-dominant. The lower panels display the $\gamma$-ray surface brightness map (${\rm SB_{\gamma}}$).  ${\rm SB_{\gamma}}$ due to hadronic and leptonic interactions are obtained from
\begin{equation} \label{eq:SBg}
\frac{\rm SB_{\rm \gamma}}{\rm erg\,s^{-1}\,cm^{-2}}= 
\left\{
	\begin{array}{ll}
\int_{-L}^{+L} dz \left[1.1\times 10^{-16}\,n_{\rm N}\, e_{\rm cr}\right]  \\
\\
\int_{-L}^{+L} dz  \left[172\times 10^{-16} \left(\frac{r}{\rm 10pc}\right)^{-2}\, e_{\rm cr}\right],
	\end{array}
\right.
\end{equation}
respectively (see Equations (\ref{eq:Lum_gamma}) and (\ref{eq:Lgic})). The hadronic $\gamma$-ray maps (panels a$1$ and b$1$) indicate that central region of the bubble is not bright in $\gamma$-ray. In contrast, for leptonic $\gamma$-ray model (panels a$2$ and b$2$), the stellar radiation field increases the $\gamma$-ray brightness in central region. This can be a diagnostic to distinguish between the hadronic and leptonic models.
\begin{figure}
\centering
\includegraphics[height= 5.8in,width=3.2in]{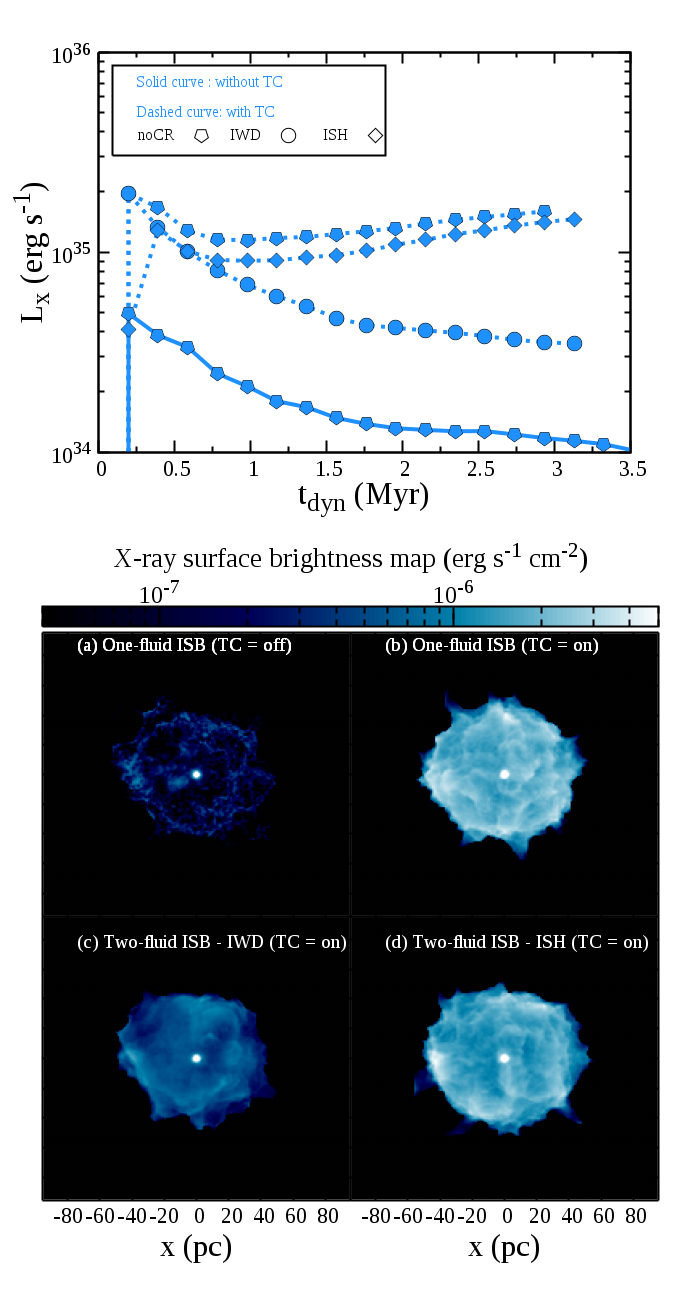}
\caption{[Top panel] Time evolution of $L_{\rm x}$ in $3$D.  The luminosity in IWD model is smaller than ISH model by a factor of $\approx 5$. [Bottom panel] X-ray ($\approx 0.5-2$ keV) surface brightness map projected on x-y plane at $2.5$ Myr.}
\label{fig:3dxray}
\end{figure}
\begin{figure}
\centering
\includegraphics[height= 2.25in,width=3.35in]{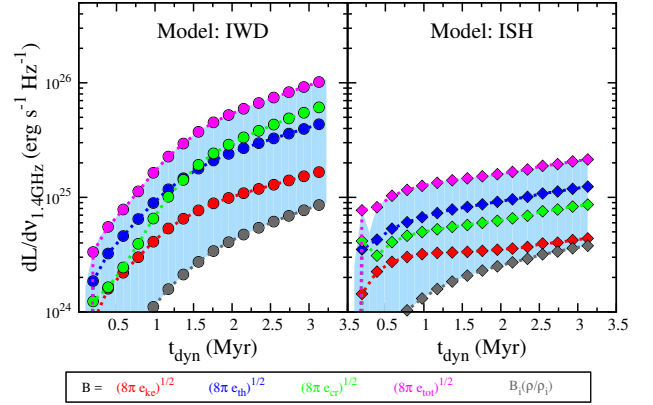}
\caption{Time evolution of radio luminosity per unit frequency for two different CR injection models in $3$D runs. }
\label{fig:3dradio}
\end{figure}

Figure \ref{fig:3dxray} for X-ray luminosity shows that the one-fluid with TC model\footnote{$L_{\rm x}$ is smaller than that of $1$D simulation. For details see Appendix \ref{sec:gridreso}.} (pentagon symbol) $L_{\rm x}\approx 1.7\times 10^{35}\, {\rm erg\,s^{-1}}$ at $t_{\rm dyn} \gtrsim 2 $ Myr. Removal of TC makes it dimmer by a factor of $\sim 10$. For ISH model, $L_{\rm x}$ approaches $\approx 1.5\times 10^{35}\, {\rm erg\,s^{-1}}$, for our choice of small $w$. $L_{\rm x}$ for the IWD model is close (difference $\lesssim 3$) to one-fluid ISB without TC. In other words, the presence of CRs can mimic the absence of thermal conduction. 

The projection maps (obtained similarly as the $\gamma$-ray map) for respective models are displayed in bottom panels. The maps clearly show that for one-fluid model, thermal conduction can increase X-ray surface brightness (as illustrated in Figure \ref{fig:illus}). Consider now the effect of CRs. If the wind-driving region (IWD) is the main site for CR acceleration, the X-ray surface brightness is dimmer than shock injection model (ISH). Therefore, the surface brightness profile is an important diagnostic to identify CR acceleration site. Later, we will show that observations prefer the ISH (brighter) model.

Figure \ref{fig:3dradio}, which displays the radio luminosity per unit frequency at $1.4$ GHz, follows a similar evolution as in $1$D runs. From this section, we conclude that the $3$D results qualitatively agree with $1$D runs.
\section{Discussions} \label{sec:discussion}
In previous section we have studied the time evolution of $\gamma$-ray, X-ray and radio luminosities, and the difference between $1$D and $3$D models. Here we explore the dependence of our results on different parameters using $1$D runs.
\subsection{Wind velocity} \label{subsec:vw}
\begin{figure}
\centering
\includegraphics[height= 5.9in,width=3.2in]{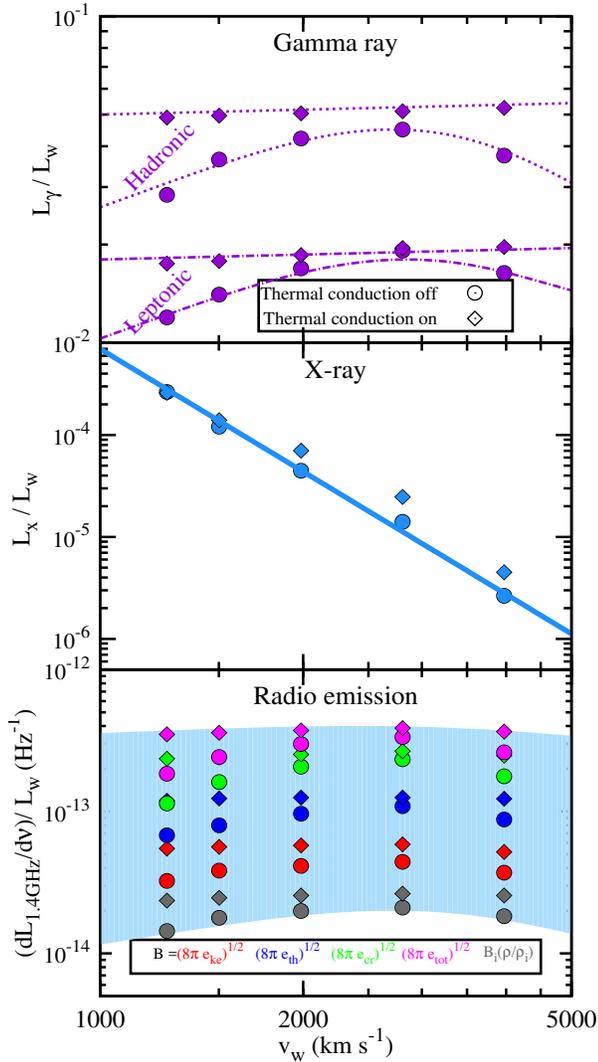}
\caption{Dependence of $\gamma$-ray, X-ray and radio luminosities (normalized w.r.t $L_{\rm w}$) on the wind velocity ($v_{\rm w}$) at $3$ Myr for IWD model. The diamond/circle symbols stand for with/without TC run. The sky-blue shaded region shows the expected range for radio luminosity.}
\label{fig:vpara}
\end{figure}

Although the wind velocity ($\approx(2L_{\rm w}/\dot{M})^{1/2}$) is a critical parameter, it does not have a well defined prescription. Figure \ref{fig:vpara} shows the dependence of our results ($L_{\rm \gamma}$, $L_{\rm x}$ and $L_{\rm R}$ in top, middle and bottom panels) on this parameter for the IWD case.
\begin{figure}
\centering
\includegraphics[height= 2.8in,width=3.2in]{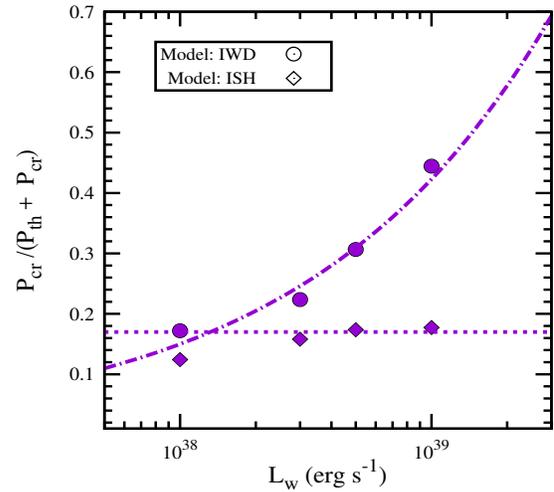}
\caption{The ratio of volume averaged cosmic ray pressure ($P_{\rm cr}$) to total pressure ($P_{\rm th} + P_{\rm cr}$) in the swept-up ISM as function of wind luminosity at $3$ Myr. We see a similar dependence for $t_{\rm dyn}\sim 1-5$ Myr.}
\label{fig:w}
\end{figure}
\begin{figure*}
\centering
\includegraphics[height= 3.5in,width=5.7in]{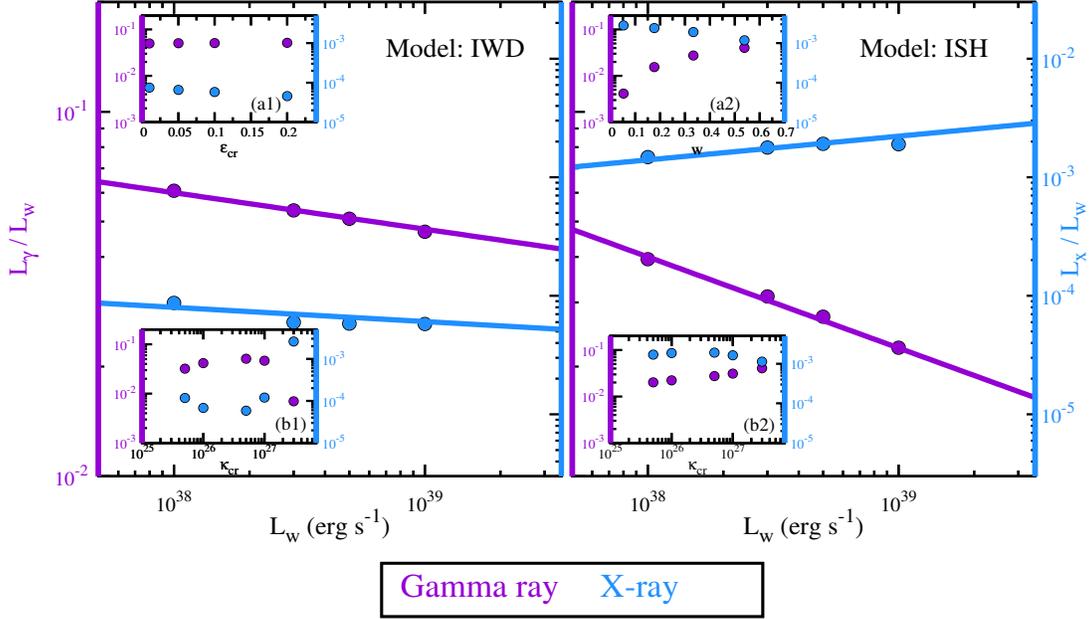}
\caption{Parametric study of $\gamma$-ray (violet) and X-ray (blue) luminosities. The violet and blue coloured vertical axes display $L_{\rm \gamma}$ and $L_{\rm x}$ respectively. The subplot (a1)/(a2) displays the dependence of results on the CR injection fraction ($\epsilon_{\rm cr}$ or $w$), and (b1)/(b2) shows the dependence on the CR diffusion coefficient. The subplot (a1) shows that $L_{\rm\gamma}$ and $L_{\rm x}$ do not depend on the injection fraction. The subplot (b1), (a2) and (b2) indicate an anti-correlation of $L_{\rm\gamma}$ with $L_{\rm x}$.}
\label{fig:para}
\end{figure*}
\begin{figure*}
\centering
\includegraphics[height= 3.5in,width=5.7in]{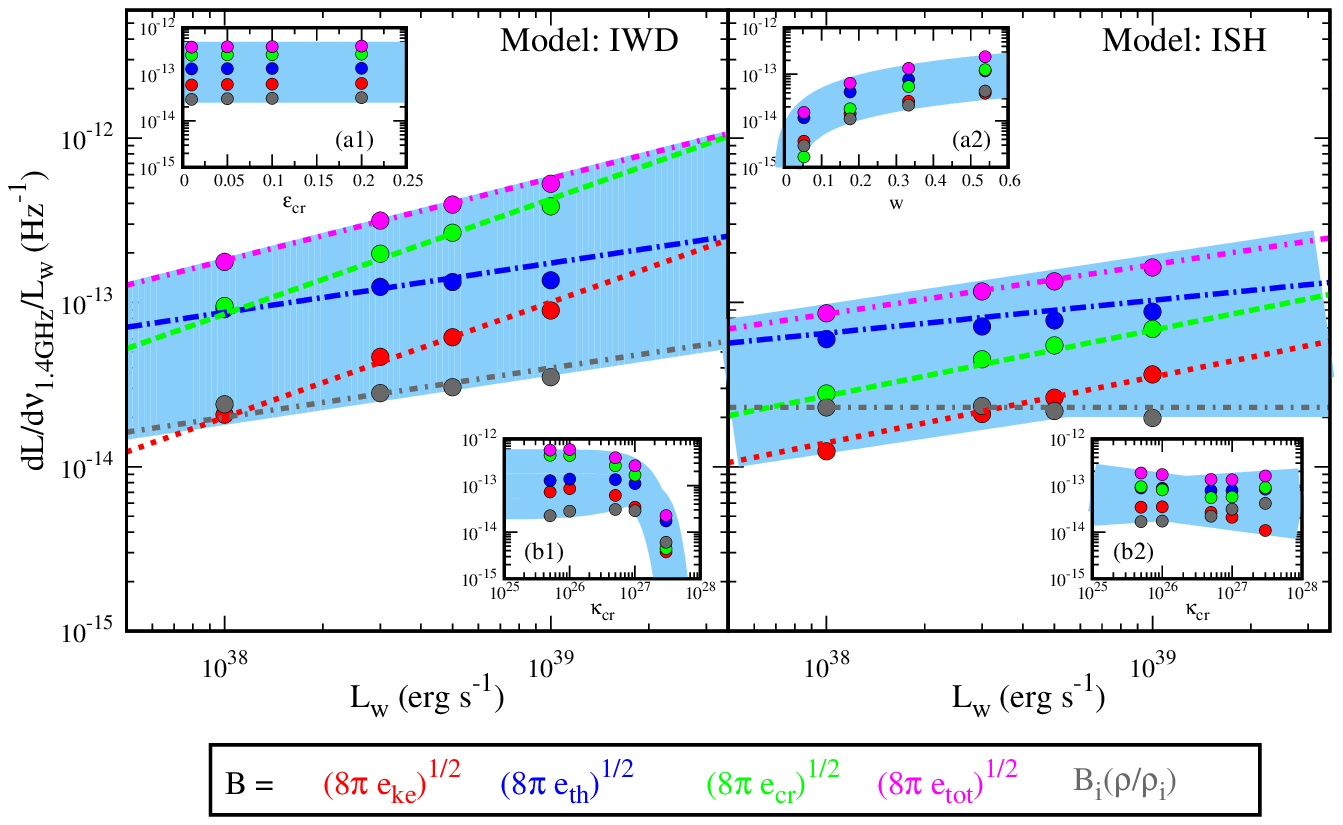}
\caption{Parametric study of the synchrotron radio luminosity per frequency ($dL/d\nu$) at $1.4$ GHz. Due to ambiguity in magnetic field ($B$), $dL/d\nu$ is spread over the sky-blue shaded region. In both panels, $L_{\rm R}$ ($=\nu dL/d\nu$) increases with $L_{\rm w}$. The subplots (a1) and (b2) display that $L_{\rm R}$ does not depend on $\epsilon_{\rm cr}$ and $\kappa_{\rm cr}$. The subplot (b1) (and (a2)) shows that $L_{\rm R}$ decreases (increases) with $\kappa_{\rm cr}$ ($w$).}
\label{fig:syn2}
\end{figure*}

The top panel of Figure \ref{fig:vpara} displays $L_{\rm \gamma}$. Two different line-styles (dotted and dash-dotted), which represent $L_{\rm \gamma}$ due to hadronic and leptonic interactions, indicate that the hadronic interaction dominates over inverse Compton scattering (section \ref{subsubsec:lepg}). When TC is off (see circles) and the wind velocity ($v_{\rm w}$) is varied from from $1000$ to $4000$ ${\rm km\,s^{-1}}$, the $\gamma$-ray luminosity changes by a factor of $\sim 2$. In contrast, for the models with TC (the diamond symbols), $L_{\rm \gamma}$ is almost independent of $v_{\rm w}$ (the violet curve). 

The middle panel displaying the X-ray luminosity shows a significant dependence on $v_{\rm w}$. A small $v_{\rm w}$ indicates a large $\dot{M}$, correspondingly a large density, and it results in a high $L_{\rm x}$. However, if $v_{\rm w}$ is too small ($v_{\rm w}\lesssim 1200\,{\rm km\,s^{-1}}$ and $\dot{M}\gtrsim 10^{-3} M{\rm _{\odot}\,yr^{-1}}$) then the shocked-wind region radiates so efficiently that it disappears and the X-ray emission is quenched ($L_{\rm x}/L_{\rm w}\ll 10^{-6}$). 

The bottom panel shows the synchrotron emission at $1.4$ GHz. This panel shows a moderate (within a factor of $\approx 2$) dependence on $v_{\rm w}$.

In ISH model, all luminosities  (not displayed) show a similar dependence on $v_{\rm w}$ when TC is off. However, in runs with TC,  the X-ray luminosity is $\gtrsim 10^{-3} L_{\rm w}$, which is significantly higher than that in IWD model.

\subsection{Star cluster mass \& CR parameters} \label{subsubsec:crpara}
Here we explore the dependence on three important parameters. The first one is the mechanical luminosity ($L_{\rm w}$) which depends on cluster mass. The other two parameters are the CR injection fraction and diffusion coefficient. In all our runs discussed in this section thermal conduction is included.

\subsubsection{Non-thermal pressure in the shell} \label{subsubsec:pcr in shell}
We have estimated the volume averaged cosmic ray and thermal pressure in the shell for four different values of $L_{\rm w}$, where the other parameters are kept identical to the fiducial run (Table \ref{tab:rundetails}). The dotted straight line in Figure \ref{fig:w} verifies that the CR pressure fraction, i.e., $W_{\rm sh}=P_{\rm cr}/(P_{\rm cr}+P_{\rm th})\approx 0.17$ is fixed in ISH. In contrast, the dash-dotted curve (IWD) shows that the CR pressure fraction increases as $W_{\rm sh}\propto L^{0.45}_{\rm w}$. This indicate that if the wind-driving region is an efficient site for CR acceleration (IWD) then for massive star clusters, CR pressure in the shell can be comparable to or larger than thermal pressure.

\subsubsection{$\gamma$-ray, X-ray and Radio} \label{subsubsec:para_gama}
In Figs. \ref{fig:para} and \ref{fig:syn2} we display the variation of $L_{\rm \gamma}$, $L_{\rm x}$ and $L_{\rm R}$ on all three parameters ($L_{\rm w}$, $w$/$\epsilon_{\rm cr}$ and $\kappa_{\rm cr}$). The main plot shows the dependence on $L_{\rm w}$, the subplots (a) and (b) show the dependence on CR injection fraction ($\epsilon_{\rm cr}$/$w$) and diffusion coefficient ($\rm \kappa_{\rm cr}$) respectively.

\begin{itemize}
\item IWD (left panels of Figures \ref{fig:para} and \ref{fig:syn2}): We find that $L_{\rm \gamma} \propto L^{0.9}_{\rm w}$, $L_{\rm x} \propto L^{0.9}_{\rm w}$ and $L_{\rm R}=\nu dL_{\rm R}/d\nu \propto L^{1.3-1.7}_{\rm w}$. The small variation of $L_{\gamma}$ can be understood from Equation (\ref{eq:Lum_g_f}) which yields $L_{\rm \gamma}\propto W_{\rm sh}\,L^{3/4}_{\rm w}$. Since $W_{\rm sh}\propto L^{0.45}_{\rm w}$ (see Figure \ref{fig:w}), we expect $L_{\rm \gamma}\propto L^{1.2}_{\rm w}$. However, in simulation we get a weaker dependence because for a low $L_{\rm  w}$ (i.e. a smaller bubble and high density ambient medium), the ambient contribution enhances $L_{\rm \gamma}$. Important point to note is that $L_{\rm x}<L_{\rm w}/10^{4}$ (blue). The subplot (a1) in both figures indicates that all luminosities are insensitive to the CR injection fraction ($\epsilon_{\rm cr}$). The subplot (b1) shows that $L_{\rm x}$ and $L_{\rm \gamma}$  are anti-correlated when $\kappa_{\rm cr}$  is varied $5\times 10^{25}\lesssim \kappa_{\rm cr}/{\rm cm^{2}\,s^{-1}}\lesssim 3\times 10^{27}$. This is because a sufficiently large (or sufficiently small) $\kappa_{\rm cr}$ diminishes the efficiency of CR re-acceleration (for details, see section $4.2$ in \citealt{Gupta2018}) which increases $L_{\rm x}$ but decreases $L_{\rm \gamma}$. We conclude that $L_{\rm \gamma}$ can be as large as $\approx 5\%$ of $L_{\rm w}$ only when $10^{26}\lesssim\kappa_{\rm cr}/{\rm cm^{2}\,s^{-1}}\lesssim 10^{27}$.

\item ISH (right panels of Figures \ref{fig:para} and \ref{fig:syn2}): Figures show that $L_{\rm \gamma} \propto L^{0.75}_{\rm w}$, $L_{\rm x} \propto L^{1.2}_{\rm w}$ and $dL_{\rm R}/d\nu \propto L^{1-1.4}_{\rm w}$.  The dependence of $L_{\rm \gamma}$ and $L_{\rm x}$ are therefore consistent with Equations (\ref{eq:Lum_g_f}) and (\ref{eq:Lxsim}). The subplot (a2) confirms that $L_{\rm \gamma}$, $L_{\rm x}$ and $L_{\rm R}$ change almost linearly with the injection fraction ($w$). The subplot (b2) shows that $L_{\rm \gamma}$ increases with $\kappa_{\rm cr}$. This is because we have used a fixed $w$ irrespective of $\kappa_{\rm cr}$. A larger CR diffusion enhances the ambient contribution and hence it increases $L_{\rm \gamma}$. In contrast, $L_{\rm x}$ is changed only by a factor $\lesssim 2$ because in this model the interior of the ISB does not depend on $\kappa_{\rm cr}$. 
\end{itemize}

\subsection{Comparison with Observation} \label{subsec:compobs}
We are now at a stage to compare with observation. Young star clusters ($\lesssim 3.5$ Myr) are powered mainly by stellar winds (\citealt{Leitherer99}, see also Figure $1$ in \citealt{Gupta2016}). If stars are distributed in a compact region then a coherent reverse (termination) shock is expected to form. Even if there are supernovae and massive transient winds from within the star cluster, we do not expect the scenario to change significantly as long as the energy deposited by the smooth winds dominates.

For $30$ Doradus, most of the massive stars are located at the central few pc region (e.g. \citealt{Massey1998}; \citealt{Selman1999}). Table \ref{tab:obs_data} shows that for $30$ Doradus $L_{\rm\gamma}/L_{\rm w}\sim 10^{-2}$, $L_{\rm x}/L_{\rm w}\sim 2\times 10^{-3}$ and $(dL_{\rm R}/d\nu)/L_{\rm w}\sim 8\times 10^{-14}\, {\rm Hz^{-1}}$. This suggests that the forward and reverse shock injection model (ISH) is the most preferable one (see the right panels in Figures \ref{fig:para} and \ref{fig:syn2} with $L_{\rm w}\approx 2\times 10^{39}\, {\rm erg\,s^{-1}}$). We can also put an upper limit of $\lesssim 0.2$ on the ratio of CR pressure to thermal pressure in the photo-ionized shell (Figure \ref{fig:w}). Moreover, our results also suggest that thermal conduction is indeed required to explain $L_{\rm x}$, because without it $L_{\rm x}/L_{\rm w}\lesssim 10^{-4}$.

For other objects listed in Table \ref{tab:obs_data}, $L_{\rm\gamma}/L_{\rm w} \lesssim 10^{-3}$. The reason could be a low density ISM. However, their structures are quite irregular, and our simplified model may not be suitable for a meaningful comparison.
\section{Conclusions}\label{sec:summary}
We have presented the detailed diagnostics of the $\gamma$-ray, X-ray and radio luminosities to understand the effects of CR in a young star cluster. This work is an extension of \citet{Gupta2018} which demonstrated the two-fluid model of an ISB. Our key results are:
\begin{enumerate}
\item Ambient medium: We have modeled an ambient density profile (mean surface density $
\sim 50$ $M{\rm_{\odot}\,pc^{-2}}$) that follows $M_{\rm cl}\propto R_{\rm cl}^{2}$ (\S \ref{subsec:ambient}, Figure \ref{fig:ambprof}). This profile makes the resulting $\gamma$-ray luminosity weakly dependent on time (\S\ref{subsubsec:hg}), and allows a convenient comparison with observation.
\item ISB profiles: The structure of ISB plays a crucial role in the comparison with observations. We have focused on the early phases of bubble evolution ($\lesssim 4$ Myr; i.e. ISB is driven by the stellar wind, not supernovae), and show that $3$D structure is consistent with $1$D runs (Figure \ref{fig:3dprofiles}).
\item Dynamical effects of CRs: We compare bubble radius between with and without CR models by considering two different CR injection scenarios (models: IWD and ISH, \S \ref{subsec:crinject}). Our models do not show a noticeable difference in the bubble radius (Figure \ref{fig:bubble_size}).
\item Multi-wavelength luminosities: We find that if central wind-driving (IWD) region accelerates CRs then $\gamma$-ray luminosity ($L_{\rm \gamma}$) can reach $\approx 5\%$ of the wind mechanical power ($L_{\rm w}$) when the reverse (termination) shock is CR dominated. In this scenario, $L_{\rm \gamma}$ is almost independent of the CR injection fraction (Figure \ref{fig:para}) and the X-ray luminosity $\lesssim L_{\rm w}/10^4$. If the forward/reverse shock (ISH) of an ISB is the CR injection site, then the $\gamma$-ray luminosity is directly proportional to injection fraction ($w$) and X-ray luminosity ($\gtrsim L_{\rm w}/10^3 $) is reduced by a factor of $\sim (1-w)^2$ from one-fluid model (Figure \ref{fig:para}). We also show the expected range for the radio emission at $1.4$ GHz (Figure \ref{fig:syn2}).
\item Comparison with observation: We compare our models with the well observed star cluster, $30$ Doradus. We find that the CR injection at the reverse and forward shocks (ISH model) can explain multi-wavelength observations.
\end{enumerate}

Therefore, we suggest that the comparison of the $\gamma$-ray, X-ray and radio luminosities with the wind mechanical power will help to know the details of CR acceleration in star clusters.

\section*{Acknowledgements}
We thank David Eichler for valuable suggestions. We are thankful to the Supercomputing Education and Research Centre (SERC), IISc for providing us the facility of Cray XC40-SahasraT cluster to perform our simulations. PS acknowledges the partial support from an India-Israel joint research grant (6-10/2014[IC]). SG acknowledges CSIR -- SPM fellowship, India for financial support.

\appendix
\section{Numerical convergence} \label{sec:gridreso}
\begin{figure*}
\centering
\includegraphics[height= 2.5in,width=6.in]{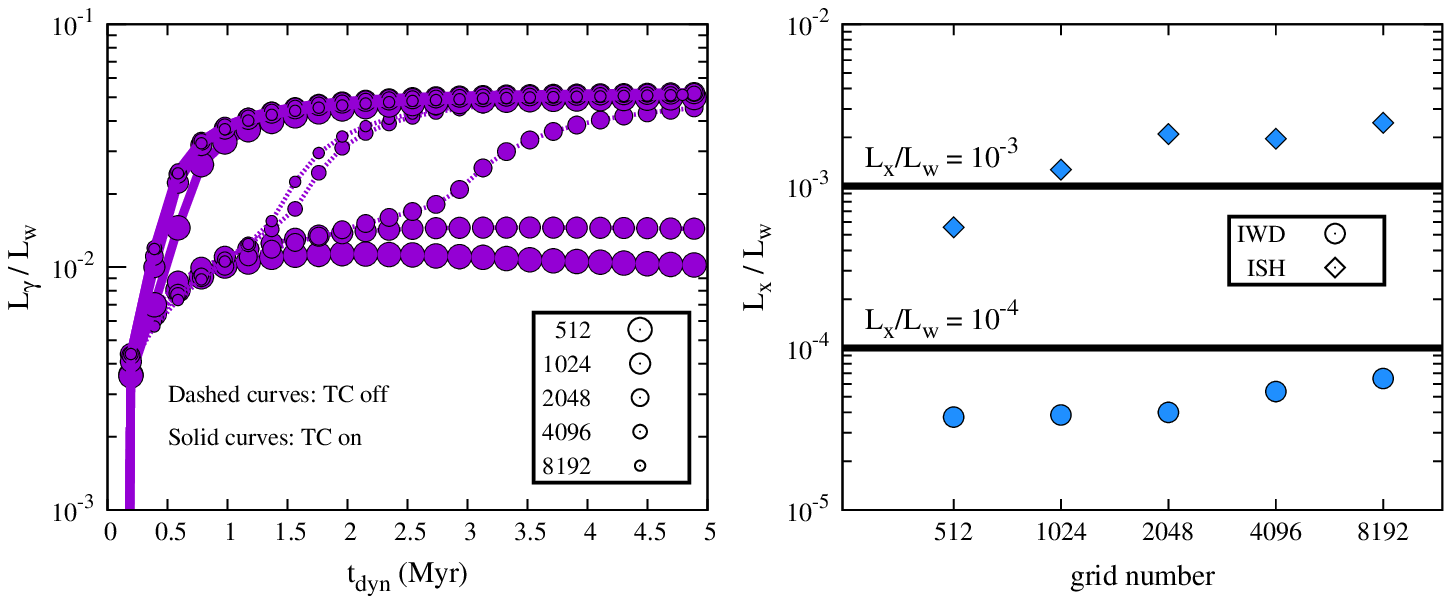}
\caption{Resolution study for our $1$D fiducial model. [Left panel] The time evolution of $L_{\rm \gamma}$ in IWD model for five different grid resolutions (fiducial resolution $\Delta r \simeq 0.061$ pc i.e. $n_{\rm grid}=4096$) where the size of the circle is proportional to grid spacing ($\Delta r$).  Figure shows that for a low resolution run (i.e., $\Delta r> 0.061$ pc), $L_{\rm \gamma}$ is converged if thermal conduction (TC) is on. [Right panel] Dependence of $L_{\rm x}$ (with TC) on grid number ($\propto 1/\Delta r$). The circle and diamond symbols stand for IWD and ISH models respectively at $3$ Myr. This figure shows that for IWD model $L_{\rm x}/L_{\rm w}\lesssim 10^{-4}$.}
\label{fig:reso}
\end{figure*}
We present resolution study for our $1$D fiducial model in Figure \ref{fig:reso}. The dashed curves (IWD model without TC) show that $L_{\rm \gamma}$ is converged if the grid resolution $\Delta r\lesssim 0.061$ pc (i.e. grid number $\gtrsim 4096$). The difference in $\gamma$-ray luminosity between low ($\Delta r\approx 0.5$ pc) and high ($\Delta r=0.03$ pc) resolution runs is a factor of $\approx 4$. For the runs with TC (solid curves), $L_{\rm \gamma}$ is almost independent of grid resolution. The difference between with and without TC in the high resolution case is indistinguishable. It suggests including thermal conduction while studying the two-fluid model. 

The right panel shows the X-ray luminosity (with TC) for five different resolutions. For ISH model, a low resolution run causes large cooling losses (see e.g. \citealt{Yadav2017}) resulting a smaller $L_{\rm x}$. In IWD model, the bubble is CR dominated which does not cool as efficiently as thermal fluid, results a weaker dependence on $\Delta r$ than the ISH case. This figure shows that for IWD model $L_{\rm x}/L_{\rm w}\lesssim 10^{-4}$.

For our $3$D runs, $\Delta r\approx0.79$ pc (for $|x, y, z|\geq 5$ pc) which is much larger than the spatial resolution used in our $1$D simulations ($\Delta r\approx0.06$ pc). We have included thermal conduction in order to get a numerically converged $L_{\gamma}$. However, $L_{\rm x}$ is underestimated at this resolution because of the reason discussed above. The qualitative results of $3$D runs are consistent with $1$D runs.

\begin{thebibliography}{}
%
\bibitem[\protect\citeauthoryear{Abdo \etal}{2010}]{Abdo2010}
Abdo, A. A., Ackermann, M., Ajello, M., Atwood, W. B. \etal  2010
\ufhref[blue]{https://doi.org/10.1051/0004-6361/200913474}{A\& A}, 
\ufhref[blue]{http://adsabs.harvard.edu/abs/2010A\%26A...512A...7A}{512, A7}
%
\bibitem[\protect\citeauthoryear{Abramowski \etal}{2015}]{Abramowski2015}
Abramowski, A., Aharonian, F., Ait Benkhali, F., Akhperjanian, A. G.\etal  2015
\ufhref[blue]{http://dx.doi.org/10.1126/science.1261313}{Science}, 
\ufhref[blue]{http://adsabs.harvard.edu/abs/2015Sci...347..406H}{347, 406}
%
\bibitem[\protect\citeauthoryear{Ackermann \etal}{2011}]{Ackermann2011} 
Ackermann, M., Ajello, M., Allafort, A., Baldini, L., \etal  2011 
\ufhref[blue]{http://adsabs.harvard.edu/abs/2011Sci...334.1103A}{Science}, 
\ufhref[blue]{http://dx.doi.org/10.1126/science.1210311}{334, 1103}
%
\bibitem[\protect\citeauthoryear{Aharonian, Yang \& de O\~{n}a Wilhelmi}{2018}]{Aharonian2018} 
Aharonian, F., Yang, R.-z. \& de O\~{n}a Wilhelmi, E. 2018
\ufhref[blue]{https://arxiv.org/abs/1804.02331}{arXiv:1804.02331}
%
\bibitem[\protect\citeauthoryear{Alexiades \etal}{1996}]{Alexiades1996}
Alexiades, V., Amiez, G., \& Gremaud P.-A. 1996,
 \ufhref[blue]{https://doi.org/10.1002/(SICI)1099-0887(199601)12:1\%3C31::AID-CNM950\%3E3.0.CO;2-5}{Com. Num. Meth. Eng.},
 \ufhref[blue]{https://doi.org/10.1002/(SICI)1099-0887(199601)12:1\%3C31::AID-CNM950\%3E3.0.CO;2-5}{12, 31}
%
\bibitem[\protect\citeauthoryear{Becker \& Kazanas}{2001}]{Becker2001}
Becker, P. A. \& Kazanas, D. 2001
\ufhref[blue]{http://dx.doi.org/10.1086/318257}{ApJ}, 
\ufhref[blue]{http://adsabs.harvard.edu/abs/2001ApJ...546..429B}{546, 429}
%
\bibitem[\protect\citeauthoryear{Booth \etal }{2013}]{Booth2013}
Booth, C. M., Agertz, O., Kravtsov, A. V., Gnedin, N. Y. 2013,
\ufhref[blue]{https://dx.doi.org/10.1088/2041-8205/777/1/L16}{ApJL}, 
\ufhref[blue]{http://adsabs.harvard.edu/abs/2013ApJ...777L..16B}{777, L16}
%
\bibitem[\protect\citeauthoryear{Bykov}{2014}]{Bykov2014}
Bykov, A. M. 2014
\ufhref[blue]{https://doi.org/10.1007/s00159-014-0077-8}{A \& ARv}, 
\ufhref[blue]{http://adsabs.harvard.edu/abs/2014A\%26ARv..22...77B}{22, 77}
\bibitem[\protect\citeauthoryear{Chevalier}{1983}]{Chevalier1983}
Chevalier, R. A. 1983,
 \ufhref[blue]{http://dx.doi.org/10.1086/161338}{ApJ}, 
 \ufhref[blue]{http://adsabs.harvard.edu/abs/1983ApJ...272..765C}{272,765}
%
\bibitem[\protect\citeauthoryear{Chevalier \&  Clegg}{1985}]{CC1985}
Chevalier, R. A. \& Clegg, A. W. 1985,
 \ufhref[blue]{http://dx.doi.org/10.1038/317044a0}{Nature}, 
 \ufhref[blue]{http://adsabs.harvard.edu/abs/1985Natur.317...44C}{317, 44}
%
\bibitem[\protect\citeauthoryear{Chu \etal}{2003}]{Chu2003a}
Chu, Y.-H., Guerrero, M. A., Gruendl, R. A., Garcia-Segura, G., Wendker, H. J. 2003,
 \ufhref[blue]{http://dx.doi.org/10.1086/379607}{ApJ}, 
 \ufhref[blue]{http://adsabs.harvard.edu/abs/2003ApJ...599.1189C}{599, 1189}
 %
 \bibitem[\protect\citeauthoryear{Crowther \& Dessart}{1998}]{Crowther1998}
Crowther P. A. \& Dessart L. 1998 
 \ufhref[blue]{http://dx.doi.org/10.1046/j.1365-8711.1998.01400.x}{MNRAS}, 
 \ufhref[blue]{http://adsabs.harvard.edu/abs/1998MNRAS.296..622C}{296, 622}
\bibitem[\protect\citeauthoryear{Dale \etal} {2013}]{Dale2013}
Dale, J. E., Ercolano, B., Bonnell, I. A. 2013,
 \ufhref[blue]{http://dx.doi.org/10.1093/mnras/sts592}{MNRAS}, 
 \ufhref[blue]{http://adsabs.harvard.edu/abs/2013MNRAS.430..234D}{430, 234}
 %
 \bibitem[\protect\citeauthoryear{Dermer}{1986}]{Dermer1986}
Dermer, C. D. 1986
 \ufhref[blue]{http://adsabs.harvard.edu/abs/1986A\%26A...157..223D}{A\& A}, 
 \ufhref[blue]{http://adsabs.harvard.edu/abs/1986A\%26A...157..223D}{157, 223}
 %
 \bibitem[\protect\citeauthoryear{Drury \& V{\"o}lk}{1981}]{Drury1981}
Drury, L. O'C. \& V{\"o}lk, J. H. 1981
 \ufhref[blue]{http://dx.doi.org/10.1086/159159}{ApJ}, 
 \ufhref[blue]{http://adsabs.harvard.edu/abs/1981ApJ...248..344D}{248, 344} 
%
 \bibitem[\protect\citeauthoryear{Drury \& Falle}{1986}]{Drury1986}
Drury, L. O'C. \& Falle, S. A. E. G. 1986
 \ufhref[blue]{https://doi.org/10.1093/mnras/223.2.353}{MNRAS}, 
 \ufhref[blue]{http://adsabs.harvard.edu/abs/1986MNRAS.223..353D}{223, 353}
\bibitem[\protect\citeauthoryear{Foreman \etal} {2015}]{Foreman2015}
Foreman, G., Chu, Y.-H., Gruendl, R., Hughes, A. \etal  2015,
 \ufhref[blue]{https://doi.org/10.1088/0004-637X/808/1/44}{ApJ}, 
 \ufhref[blue]{http://adsabs.harvard.edu/abs/2015ApJ...808...44F}{808,44}
 %
 \bibitem[\protect\citeauthoryear{Guo \& Oh}{2008}]{Guo2008}
Guo, F.,  \&  Oh, S. P., 2008
 \ufhref[blue]{http://cdsads.u-strasbg.fr/abs/2008MNRAS.384..251G}{MNRAS}, 
 \ufhref[blue]{http://dx.doi.org/10.1111/j.1365-2966.2007.12692.x}{384,251}
 %
\bibitem[\protect\citeauthoryear{Gupta \etal}{2016}]{Gupta2016}
Gupta, S., Nath, B. B., Sharma, P. \& Shchekinov, Y. 2016 
 \ufhref[blue]{http://dx.doi.org/10.1093/mnras/stw1920}{MNRAS}, 
 \ufhref[blue]{http://adsabs.harvard.edu/abs/2016MNRAS.462.4532G}{462, 4532} 
%
\bibitem[\protect\citeauthoryear{Gupta \etal}{2018}]{Gupta2018}
Gupta, S., Nath, B. B., Sharma, P. \& Eichler, D.  2018
\ufhref[blue]{https://doi.org/10.1093/mnras/stx2427}{MNRAS}, 
\ufhref[blue]{http://adsabs.harvard.edu/abs/2018MNRAS.473.1537G}{473,1537}
%
\bibitem[\protect\citeauthoryear{Harayama, Eisenhauer \& Martins}{2014}]{Harayama2014}
Harayama, Y., Eisenhauer, F. \& Martins, F.  2014,
\ufhref[blue]{https://doi.org/10.1086/524650}{ApJ}, 
\ufhref[blue]{https://doi.org/10.1086/524650}{675,1319}
%
\bibitem[\protect\citeauthoryear{Harper-Clark \& Murray}{2009}]{Harper2009}
Harper-Clark, E. \& Murray, N. 2009, 
 \ufhref[blue]{http://dx.doi.org/10.1088/0004-637X/693/2/1696}{ApJ}, 
 \ufhref[blue]{http://dx.doi.org/10.1088/0004-637X/693/2/1696}{693, 1696}
 %
\bibitem[\protect\citeauthoryear{Hughes \etal}{2007}]{Hughes2004}
Hughes, A., Staveley-Smith, L., Kim, S., Wolleben, M. \& M. Filipovic, 2007
 \ufhref[blue]{http://dx.doi.org/10.1111/j.1365-2966.2007.12466.x}{MNRAS}, 
 \ufhref[blue]{http://adsabs.harvard.edu/abs/2007MNRAS.382..543H}{382, 543}
 %
\bibitem[\protect\citeauthoryear{Hughes \etal}{2010}]{Hughes2010}
Hughes, A., Wong, T., Ott, J.,  Muller, E. \etal 2010
\ufhref[blue]{https://doi.org/10.1111/j.1365-2966.2010.16829.x}{MNRAS}, 
\ufhref[blue]{http://adsabs.harvard.edu/abs/2010MNRAS.406.2065H}{406, 2065}
\bibitem[\protect\citeauthoryear{Kim \& Ostriker}{2015}]{Kim2015}
Kim, C.-G. \&  Ostriker, E. C. 2015
\ufhref[blue]{https://doi.org/10.1088/0004-637X/802/2/99}{ApJ},
 \ufhref[blue]{http://adsabs.harvard.edu/abs/2015ApJ...802...99K}{802, 99}
%
\bibitem[\protect\citeauthoryear{Kim \etal}{2003}]{Kim2003}
Kim, S., Staveley-Smith, L., Dopita, M. A., Sault, R. J., Freeman, K. C. \etal 2003
\ufhref[blue]{https://doi.org/10.1086/376980}{ApJS},
 \ufhref[blue]{http://adsabs.harvard.edu/abs/2003ApJS..148..473K}{148, 473}
%
\bibitem[\protect\citeauthoryear{Kn{\"o}dlseder}{2013}]{Knodlseder2013}
Kn{\"o}dlseder, J. 2013
\ufhref[blue]{https://doi.org/10.1007/978-3-642-35410-6_13}{ASSP},
 \ufhref[blue]{http://adsabs.harvard.edu/abs/2013ASSP...34..169K}{34, 169}
 %
\bibitem[\protect\citeauthoryear{Krause \etal}{2013}]{Krause2013}
Krause, M., Fierlinger, K., Diehl, R., Burkert, A., Voss, R., J. \etal 2013
\ufhref[blue]{https://doi.org/10.1051/0004-6361/201220060}{A\& A},
 \ufhref[blue]{http://adsabs.harvard.edu/abs/2013A\%26A...550A..49K}{550, 49}
\bibitem[\protect\citeauthoryear{Leitherer \etal}{1999}]{Leitherer99}
Leitherer, C., Schaerer, D., Goldader, J. D., et al. 1999,
 \ufhref[blue]{http://dx.doi.org/10.1086/313233}{ApJS},
 \ufhref[blue]{http://adsabs.harvard.edu/abs/1999ApJS..123....3L}{123, 3}
 %
 \bibitem[\protect\citeauthoryear{Longair}{2011}]{Longair2011}
Longair M. S., 2011, High energy astrophysics,
 \ufhref[blue]{http://adsabs.harvard.edu/abs/2011hea..book.....L}{Cambridge University Press},
 third edition, Cambridge, NY 
%
\bibitem[\protect\citeauthoryear{Longmore \etal}{2014}]{Longmore2014}
Longmore, S. N., Kruijssen, J. M. D., Bastian, N., Bally, J., \etal 2014
\ufhref[blue]{http://dx.doi.org/10.2458/azu_uapress_9780816531240-ch013}{Protostars and Planets VI}, 
\ufhref[blue]{http://adsabs.harvard.edu/abs/2014prpl.conf..291L}{914,291}
%
\bibitem[\protect\citeauthoryear{Lopez \etal}{2014}]{Lopez2014}
Lopez, L. A., Krumholz, M. R., Bolatto, A. D., Prochaska, J. X.,  Ramirez-Ruiz, E. \& Castro, D. 2014,
 \ufhref[blue]{http://dx.doi.org/10.1088/0004-637X/795/2/121}{ApJ},
 \ufhref[blue]{http://adsabs.harvard.edu/abs/2014ApJ...795..121L}{795,121}
 %
\bibitem[\protect\citeauthoryear{Mannheim \& Schlickeiser}{1994}]{Mannheim1994}
Mannheim, K. \& Schlickeiser, R.  1994
\ufhref[blue]{http://adsabs.harvard.edu/abs/1994A\%26A...286..983M}{A\& A}, 
\ufhref[blue]{http://adsabs.harvard.edu/abs/1994A\%26A...286..983M}{286, 983}
%
\bibitem[\protect\citeauthoryear{Martizzi, Faucher-Gigu\'{e}re \& Quataert}{2015}]{Martizzi2015}
Martizzi, D., Faucher-Gigu\'{e}re, C.-A., Quataert, E. 2015
\ufhref[blue]{https://doi.org/10.1093/mnras/stv562}{MNRAS}, 
\ufhref[blue]{http://adsabs.harvard.edu/abs/2015MNRAS.450..504M}{450, 504}
%
\bibitem[\protect\citeauthoryear{Massey \& Hunter}{1998}]{Massey1998}
Massey, P. \& Hunter, D. A. 1998
\ufhref[blue]{https://doi.org/10.1086/305126}{ApJ}, 
\ufhref[blue]{http://adsabs.harvard.edu/abs/1998ApJ...493..180M}{493, 180}
%
 \bibitem[\protect\citeauthoryear{Maurin \etal} {2016}]{Maurin2016}
Maurin, G., Marcowith, A., Komin, N. \etal 2016,
 \ufhref[blue]{http://dx.doi.org/10.1051/0004-6361/201628465}{A\& A}, 
 \ufhref[blue]{http://adsabs.harvard.edu/abs/2016A\%26A...591A..71M}{591, 71}
%
\bibitem[\protect\citeauthoryear{Mignone \etal} {2007}]{Mignone2007}
Mignone, A., Bodo, G., Massaglia, S., Matsakos, T. \etal  2007,
 \ufhref[blue]{http://dx.doi.org/10.1086/513316}{ApJS}, 
 \ufhref[blue]{http://adsabs.harvard.edu/abs/2007ApJS..170..228M}{170, 228} 
 %
\bibitem[\protect\citeauthoryear{Muno \etal}{2006}]{Muno2006}
Muno, M. P., Law, C., Clark, J. S., Dougherty, S. M. \etal  2006
\ufhref[blue]{https://doi.org/10.1086/507175}{ApJ}, 
\ufhref[blue]{http://adsabs.harvard.edu/abs/2006ApJ...650..203M}{650,203}
%
\bibitem[\protect\citeauthoryear{Murphy \etal}{2012}]{Murphy2012}
Murphy, E. J., Porter, T. A., Moskalenko,  I. V., Helou, G. \& Strong, A. W.  2012
\ufhref[blue]{https://doi.org/10.1088/0004-637X/750/2/126}{ApJ}, 
\ufhref[blue]{http://adsabs.harvard.edu/abs/2012ApJ...750..126M}{750, 126}
%
%
\bibitem[\protect\citeauthoryear{Natarajan \& Lynden-Bell}{1997}]{Natarajan1997}
Natarajan, P. \& Lynden-Bell, D. 1997
\ufhref[blue]{https://doi.org/10.1093/mnras/286.2.268}{MNRAS}, 
\ufhref[blue]{http://adsabs.harvard.edu/abs/1997MNRAS.286..268N}{286, 268}
%
\bibitem[\protect\citeauthoryear{Nath \& Shchekinov}{2013}]{Nath2013}
 Nath, B. B. \& Shchekinov, Y. 2013
 \ufhref[blue]{https://doi.org/10.1088/2041-8205/777/1/L12}{ApJL}, 
 \ufhref[blue]{http://adsabs.harvard.edu/abs/2013ApJ...777L..12N}{777, 1}
 %
\bibitem[\protect\citeauthoryear{Ohm, Hinton \& White}{2013}]{Ohm2013}
Ohm, S., Hinton, J. A., White, R. 2013
\ufhref[blue]{http://dx.doi.org/10.1093/mnras/stt1170}{MNRAS}, 
\ufhref[blue]{http://adsabs.harvard.edu/abs/2013MNRAS.434.2289O}{434, 2289}
%
\bibitem[\protect\citeauthoryear{Persic \& Rephaeli}{2014}]{Persic2014}
Persic, M., Rephaeli, Y. 2014
\ufhref[blue]{https://arxiv.org/abs/1405.3107}{arxiv: 1405.3107}, 
\ufhref[blue]{http://adsabs.harvard.edu/abs/2015mgm..conf.1036P}{3, 1036}
%
\bibitem[\protect\citeauthoryear{Pfalzner}{2016}]{Pfalzner2015}
Pfalzner, S., Kirk, H., Sills, A., Urquhart, J. S., Kauffmann, J. \etal 2016
\ufhref[blue]{https://doi.org/10.1051/0004-6361/201527449}{A\& A}, 
\ufhref[blue]{http://adsabs.harvard.edu/abs/2016A\%26A...586A..68P}{586, 68}
%
\bibitem[\protect\citeauthoryear{Pfrommer \&  En$\upbeta$lin}{2004}]{Pfrommer2004}
Pfrommer, C. \& En$\upbeta$lin, T. A. 2004
\ufhref[blue]{https://doi.org/10.1051/0004-6361:20031464}{A\& A}, 
\ufhref[blue]{http://adsabs.harvard.edu/abs/2004A\%26A...413...17P}{413, 17}
%
\bibitem[\protect\citeauthoryear{Pfrommer \etal}{2017}]{Pfrommer2017}
Pfrommer, C., Pakmor, R., Schaal, K., Simpson, C. M. \etal 2017
\ufhref[blue]{https://doi.org/10.1093/mnras/stw2941}{MNRAS}, 
\ufhref[blue]{http://adsabs.harvard.edu/abs/2017MNRAS.465.4500P}{465, 4500}
%
\bibitem[\protect\citeauthoryear{Rosen \etal }{2014}]{Rosen2014}
Rosen, A. R., Lopez, L. A, Krumholz M. R., Ramirez-Ruiz, E. 2014 
\ufhref[blue]{http:/dx.doi.org/10.1093/mnras/stu1037}{MNRAS}, 
 \ufhref[blue]{http://adsabs.harvard.edu/abs/2014MNRAS.442.2701R}{442, 2701}
 %
\bibitem[\protect\citeauthoryear{Rybicki \& Lightman}{1979}]{Rybicki1979}
Rybicki G. B. \& Lightman A. P. 1979 Radiative processes in astrophysics
\ufhref[blue]{http://adsabs.harvard.edu/abs/1979AstQ....3..199R}{Wiley}
%
\bibitem[\protect\citeauthoryear{Salem \& Bryan}{2014}]{Salem2013}
Salem, M \& Bryan, G. L. 2014,
 \ufhref[blue]{https://doi.org/10.1093/mnras/stt2121}{MNRAS}, 
 \ufhref[blue]{http://adsabs.harvard.edu/abs/2014MNRAS.437.3312S}{437, 3312}
 %
 \bibitem[\protect\citeauthoryear{Selman \etal}{1999}]{Selman1999}
Selman, E.,  Melnick, J.,  Bosch, G., \& Terlevich, R. 1999,
 \ufhref[blue]{http://adsabs.harvard.edu/abs/1999A\%26A...347..532S}{A\&A}, 
 \ufhref[blue]{http://adsabs.harvard.edu/abs/1999A\%26A...347..532S}{347, 532}
 %
\bibitem[\protect\citeauthoryear{Sharma \etal}{2014}]{Sharma2014}
Sharma, P., Roy, A., Nath, B. B., Shchekinov, Y. 2014,
 \ufhref[blue]{http://dx.doi.org/10.1093/mnras/stu1307}{MNRAS}, 
 \ufhref[blue]{http://adsabs.harvard.edu/abs/2014MNRAS.443.3463S}{443, 3463}
%
 \bibitem[\protect\citeauthoryear{Silich \& Tenorio-Tagle}{2013}]{Silich2013}
Silich, S., Tenorio-Tagle, G., 2013,
 \ufhref[blue]{http://dx.doi.org/10.1088/0004-637X/765/1/43}{ApJ},
 \ufhref[blue]{http://adsabs.harvard.edu/abs/2013ApJ...765...43S}{765, 43}
 %
\bibitem[\protect\citeauthoryear{Solomon \etal}{1987}]{Solomon1987}
Solomon, P. M., Rivolo, A. R., Barrett, J.\& Yahil, A. 1987
\ufhref[blue]{http://adsabs.harvard.edu/abs/1987ApJ...319..730S}{ApJ}, 
\ufhref[blue]{http://adsabs.harvard.edu/abs/1987ApJ...319..730S}{319, 730}
%
\bibitem[\protect\citeauthoryear{Simpson \etal}{2016}]{Simpson2016}
Simpson, C. M., Pakmor, R., Marinacci, F., Pfrommer, C., Springel, V. \etal 2016,
 \ufhref[blue]{https://doi.org/10.3847/2041-8205/827/2/L29}{ApJ}, 
 \ufhref[blue]{http://adsabs.harvard.edu/abs/2016ApJ...827L..29S}{827,29}
\bibitem[\protect\citeauthoryear{Tibaldoa}{2013}]{Tibaldoa2013}
Tibaldo, L., Grenierb, I. A. Fermi LAT collaboration 2013
\ufhref[blue]{https://doi.org/10.1016/j.nuclphysbps.2013.05.011}{Nuclear Physics B (Proc. Suppl.)}, 
\ufhref[blue]{https://doi.org/10.1016/j.nuclphysbps.2013.05.011}{239, 70}
%
 \bibitem[\protect\citeauthoryear{Townsley \etal}{2006}]{Townsley2006}
Townsley, L. K., Broos, P. S., Feigelson, E. D., Brandl, B. R. \etal 2006,
 \ufhref[blue]{http://dx.doi.org/10.1086/500532}{ApJ}, 
 \ufhref[blue]{http://adsabs.harvard.edu/abs/2006AJ....131.2140T}{131, 2140} 
 \bibitem[\protect\citeauthoryear{Valle}{1993}]{Valle1993}
Valle, J. P. 1993
 \ufhref[blue]{http://dx.doi.org/10.1086/173517}{ApJ}, 
 \ufhref[blue]{http://adsabs.harvard.edu/abs/1993ApJ...419..670V}{419, 670} 
 %
 \bibitem[\protect\citeauthoryear{Vasiliev, Shchekinov \& Nath}{2017}]{Vasiliev2017}
Vasiliev, E. O., Shchekinov, Y. \& Nath, B. B. 2017
\ufhref[blue]{https://doi.org/10.1093/mnras/stx719}{MNRAS}
 \ufhref[blue]{http://adsabs.harvard.edu/abs/2017MNRAS.468.2757V}{468,2757}
 %
%
\bibitem[\protect\citeauthoryear{Weaver \etal}{1977}]{Weaver1977}
Weaver, R., McCray, R., Castor, J., Shapiro, P., Moore, R., 1977,
 \ufhref[blue]{http://dx.doi.org/10.1086/155692}{ApJ},
 \ufhref[blue]{http://adsabs.harvard.edu/abs/1977ApJ...218..377W}{218, 377}
 %
  \bibitem[\protect\citeauthoryear {Wiener, Pfrommer \& Oh}{2017}]{Wiener2017}
Wiener, J., Pfrommer, C. \& Oh, S. P. 2017, 
 \ufhref[blue]{http://dx.doi.org/10.1093/mnras/stx127}{MNRAS}, 
 \ufhref[blue]{http://adsabs.harvard.edu/abs/2017MNRAS.467..906W}{467, 906}
 %
\bibitem[\protect\citeauthoryear{ Wright \etal}{2010}]{Wright2010}
Wright, N. J., Drake, J. J., Drew, J. E., \& Vink, J. S. 2010
\ufhref[blue]{http://dx.doi.org/10.1088/0004-637X/713/2/871}{ApJ},
\ufhref[blue]{http://adsabs.harvard.edu/abs/2010ApJ...713..871W}{713, 871}  
%
\bibitem[\protect\citeauthoryear{Yadav \etal}{2017}]{Yadav2017}
Yadav, N., Mukherjee, D., Sharma, P., Nath, B. B. 2017,
 \ufhref[blue]{http://dx.doi.org/10.1093/mnras/stw2522}{MNRAS}
 \ufhref[blue]{http://adsabs.harvard.edu/abs/2017MNRAS.465.1720Y}{465, 172}
 %
\bibitem[\protect\citeauthoryear{Yang \& Aharonian}{2017}]{Yang2017}
Yang, R.-z. \& Aharonian, F. 2017
 \ufhref[blue]{http://dx.doi.org/10.1051/0004-6361/201630213}{A \& A},
 \ufhref[blue]{http://adsabs.harvard.edu/abs/2017A\%26A...600A.107Y}{600, A107}
 %
 \bibitem[\protect\citeauthoryear{Yang, de O\~{n}a Wilhelmi \& Aharonian}{2018}]{Yang2018}
Yang, R.-z., de O\~{n}a Wilhelmi, E., Aharonian, F. 2018
 \ufhref[blue]{https://doi.org/10.1051/0004-6361/201732045}{A \& A},
 \ufhref[blue]{http://adsabs.harvard.edu/abs/2018A\%26A...611A..77Y}{611, A77}
 %
\end{thebibliography}
\end{document}